\pdfoutput=1
\documentclass[aps,prd,reprint,showpacs,preprintnumbers,superscriptaddress,
nofootinbib,eqsecnum]{revtex4-1}

\usepackage{enumerate}
\usepackage[utf8]{inputenc}
\usepackage{rotating}
\usepackage{amsmath}
\usepackage{amssymb}
\usepackage{amscd}
\usepackage{mathrsfs}
\usepackage{graphicx}
\usepackage{epstopdf}
\usepackage{setspace}
\usepackage{fancyheadings}
\usepackage{multirow}
\usepackage{cancel}
\usepackage{array}
\usepackage{natbib}
\usepackage[colorlinks]{hyperref}
\hypersetup{
pdfauthor={Nicole F. Bell, Yi Cai and Anibal Medina},
pdftitle={Co-annihilating Dark Matter: Effective Operator Analysis and Collider Phenomenology}}

\allowdisplaybreaks[1]

\begin{document}

\title{Co-annihilating Dark Matter: Effective Operator Analysis and Collider Phenomenology}
\author{Nicole F. Bell}\email{n.bell@unimelb.edu.au}
\affiliation{ARC Centre of Excellence for Particle Physics at the Terascale, School of Physics, The University of Melbourne, Victoria 3010, Australia}
\author{Yi Cai}\email{yi.cai@unimelb.edu.au}
\affiliation{ARC Centre of Excellence for Particle Physics at the Terascale, School of Physics, The University of Melbourne, Victoria 3010, Australia}
\author{Anibal D. Medina}\email{amedina@physics.unimelb.edu.au}
\affiliation{ARC Centre of Excellence for Particle Physics at the Terascale, School of Physics, The University of Melbourne, Victoria 3010, Australia}

\begin{abstract}
We study dark matter (DM) models in which there are two dark sector particles, $\chi_1$ and $\chi_2$, of near mass.  In such models, co-annihilation of $\chi_1$ and $\chi_2$ may be the dominant process controlling the DM relic density during freezeout in the early universe.  In this scenario, there is no significant contribution to direct and indirect detection signals, unless there exists an extreme degeneracy in the masses of the lightest dark sector particles.  Therefore, relic density constraints and collider searches provide the most relevant information about these models.  We consider Dirac fermion dark matter which couples to standard model (SM) particles via an effective operator.  For the collider phenomenology, where an effective field theory may not be valid, we adopt a simple Z' model to provide an appropriate UV completion.  We explore the interesting LHC signals that arise from the dark matter production process $pp \rightarrow \overline{\chi_1} + \chi_2 + \textrm{ jet}$, followed by the decay $\chi_2 \rightarrow \chi_1 + SM$.
\end{abstract}

\maketitle


\section{Introduction}
\label{sec:Introduction}

Despite significant experimental effort to uncover the nature of dark
matter via direct detection~\cite{Agnese:2013rvf, Aalseth:2012if, Angloher:2011uu, Savage:2008er, 
Ahmed:2009zw, Ahmed:2010wy, Agnese:2013jaa, Akerib:2013tjd, Angle:2011th}, 
collider searches~\cite{ATLAS:2012zim,ATLAS:2012ky, Chatrchyan:2012tea, Chatrchyan:2012me, CMS:2013iea}, 
and indirect detection signals~\cite{Adriani:2008zr, Adriani:2010ib, Barwick:1997ig, Aguilar:2007yf, FermiLAT:2011ab,
Adriani:2008zq, Adriani:2010rc, Chang:2008aa, Aharonian:2009ah}, it still remains elusive.  The predictions
for the signals in these experiments are, in general, quite model
dependent.  However, an effective field theory (EFT) formalism is a
useful way to link results from these searches in a model independent
framework~\cite{Goodman:2010yf,Goodman:2010ku,Dreiner:2012xm,Beltran:2010ww,
Fox:2011pm}.

In the EFT approach, we use a set of non-renormalizable effective
operators to parametrise the interaction of a pair of DM particles
with standard model (SM) particles.  The EFT operators are constrained
only by Lorentz and gauge invariance, and would be obtained from a UV
complete theory by integrating out the particle that mediates the
interaction.  For example, if we assume the dark matter is fermionic, the
lowest order effective operators involving the DM, $\chi$, and SM
fermions, $f$, are of the form
\begin{equation}
\frac{1}{\Lambda_{\rm eff}^2} (\overline{\chi}\Gamma_1\chi)(\overline{f}\Gamma_2 f) , 
\end{equation}
where $\Gamma_1$ and $\Gamma_2$ are combinations of Dirac
matrices. These operators are suppressed by two powers of an effective
scale
\begin{equation}
\Lambda_{\rm eff}= \frac{M}{\sqrt{g_{SM} g_{DM}}} ,
\end{equation}
where $M$ is the mass of the heavy mediator, which couples to the
DM and SM particles with strength $g_{SM}$ and $g_{DM}$ respectively.

However, while the simple EFT approach is a useful starting point for
describing DM interactions, it may prove to be an inadequate
description.  For many choices of the effective operator, the
parameter space is now quite constrained, particularly for low values
of the dark matter mass.  There is some tension between the lower
limits on $\Lambda_{\rm eff}$ obtained from the absence of signals in
direct detection and colliders experiments, and the upper limits on
$\Lambda_{\rm eff}$ necessary to have sufficient annihilation in the
early Universe such that the relic density does not lead to
overclosure \cite{Zheng:2010js,Buckley:2011kk}. Moreover, it is
dangerous to blindly apply an EFT approach to collider
studies~\cite{Busoni:2013lha}.  Given that the effective scale
$\Lambda_{\rm eff}$ can be as light as several hundred GeV, the
momentum transfer in LHC interactions may easily exceed the mass of
the particle which mediates the interaction.

An implicit assumption in the standard EFT description is a separation
of scales which enables all the dark sector particles other than the
DM candidate itself to be integrated out.  This will break down if
there are other dark sector particles with masses comparable to the
DM.  A well motivated example is the co-annihilation scenario, in
which the dark sector contains two near-mass particles, $\chi_1$ and
$\chi_2$, whose freezeout in the early universe is
coupled~\cite{Griest:1990kh,Edsjo:1997bg}.  The relic density is
controlled by the (co-) annihilation processes,
$\overline\chi_i \chi_j \rightarrow SM$, for $i,j =\{1,2\}$, rather
than $\overline\chi_1 \chi_1$ annihilation alone.  With
$m_{\chi_2} \gtrsim m_{\chi_1}$, the $\chi_2$ all decay to the stable
DM candidate $\chi_1$, which forms the relic density in the universe
today.  However, because the relic density can be controlled by
processes involving $\chi_2$, while the direct and indirect detection
signals involve $\chi_1$ alone, the tensions between relic density and
direct/indirect detection constraints are alleviated.

For the usual case of self-annihilating dark matter, collider constraints on
$\Lambda_{\rm eff}$ are obtained from mono-jet~\cite{Cao:2009uw,Fox:2011pm,
ATLAS:2012ky,Chatrchyan:2012me,Zhou:2013fla}, mono-photon~\cite{Aad:2012fw,Chatrchyan:2012tea,
Zhou:2013fla}, 
or mono-W/Z~\cite{Bell:2012rg,Carpenter:2012rg,Bai:2012xg,Zhou:2013fla,  Aad:2013oja}
 searches for DM production.  These
signals are obtained when a single object, produced via initial state
radiation (ISR) of a gluon, photon, or electroweak gauge boson, recoils
against the missing transverse momentum attributed to dark matter.

In contrast, the co-annihilation model will have distinct collider
signals due the decay $\chi_2 \rightarrow \chi_1 + SM$.  In principle,
production of $\overline\chi_i\chi_2$ could be identified by the SM
particles produced in $\chi_2$ decay, plus missing energy carried off
by the $\chi$.  However, if the mass difference $\Delta
m_{\chi}=m_{\chi_2}-m_{\chi_1}$ is relatively low, $O$(10 GeV), the SM
particles will be soft and the missing $E_T$ low.  While this signal
could potentially be detected at a future lepton collider, it would be
hidden in QCD backgrounds at a hadronic machine such as the LHC.
We shall therefore be interested in the pair production of
$\overline{\chi_i}\chi_j$ together with ISR,
followed by $\chi_2$ decay.  The ISR ensures a sufficiently large
missing $E_T$ to allow the signal to be identified.
This leads to distinct collider signatures:

\begin{enumerate}[(i)]
\item
standard mono-jet plus $\cancel{E}_{T}$ signals result from
$\overline\chi_1\chi_1 j$ production, or from $\overline\chi_1 \chi_2
j$ and $\overline\chi_2 \chi_2 j$ where the $\chi_2$ decays invisibly
(e.g. to neutrinos) or to quarks or leptons which are too soft to
detect.
\item
jet + $\cancel{E}_{T}$ + $\overline{f}f$  results from $\overline\chi_1\chi_2 j$ production.
\item
jet + $\cancel{E}_{T}$ + $\overline{f}f$ $\overline{f'}f'$ results from $\overline\chi_2\chi_2 j$ production.
\end{enumerate}

These signals provide complementary information about the dark matter
model.  The mono-jet plus $\cancel{E}_{T}$ signal is of course not
unique to the co-annihilation scenario.  However, the observation of
signal (ii) or (iii), together with (i), could be interpreted as
evidence for co-annihilation.  We shall study process (ii) in detail.

An outline of the paper is as follows: We define EFT and UV complete
versions of our model in section~\ref{sec:EFT}, compute relic density
constraints in section~\ref{sec:relic}, and determine di-jet/di-lepton
based coupling bounds in section~\ref{sec:z'constraints}.  Finally, in
section~\ref{sec:collider}, we explore co-annihilation collider
signals at the LHC (process (ii) above).

\section{Effective field theory and UV completion}
\label{sec:EFT}

We generalize the standard EFT treatment to the case of co-annihilation by considering the following effective operators:
\begin{eqnarray}
&& \frac{1}{\Lambda_{11}^2}(\overline{\chi_1}\Gamma_1 \chi_1) (\overline{f}\Gamma_2 f)\; , 
\\ 
&& \frac{1}{\Lambda_{12}^2}(\overline{\chi_1}\Gamma_1 \chi_2 )(\overline{f}\Gamma_2 f) +h.c.\;,
\\ 
&& \frac{1}{\Lambda_{22}^2}(\overline{\chi_2}\Gamma_1 \chi_2) (\overline{f}\Gamma_2 f)\;,
\end{eqnarray}
where $\chi_1$ and $\chi_2$ are Dirac fermions.
In the examples we present below we adopt vector operators with
$\Gamma_1 = \Gamma_2 = \gamma_\mu$.  We assume throughout this work that
$\chi_1$ and $\chi_2$ have similar masses, and take $\Lambda_{11}\gg
\Lambda_{12}, \Lambda_{22}$.  This ensures the self-annihilation rate
of $\chi_1$ is subdominant, and hence the co-annihilation processes
truly control the relic density determination.  Collider production of
dark sector particles will thus be dominated by $\chi_1 \overline{\chi_2}$ and
$\chi_2 \overline{\chi_2}$ states, with a negligible rate for pair production of
the DM, $\chi_1 \overline{\chi_1}$.

In the limit that $\Lambda_{11}\gg \Lambda_{12}, \Lambda_{22}$, direct
detection signals will be highly suppressed.  Direct detection signals
of the inelastic type~\cite{TuckerSmith:2001hy,TuckerSmith:2004jv}, $\chi_1 + N
\rightarrow \chi_2 + N$, will not occur unless the mass splitting
between $\chi_1$ and $\chi_2$ is smaller than the energy transfer in a
DM-nucleus scattering interaction, $\Delta
m_{\chi}=m_{\chi_2}-m_{\chi_1}\lesssim 100$~KeV.  This is much smaller
than the mass splittings we consider. (See Ref.~\cite{Bai:2011jg} for
collider signatures of inelastic dark matter.)  
Hence, although direct
detection provides very strong constraints in the usual
self-annihilation scenario (at least for effective operators that lead
to spin independent DM-nucleon scattering) this link is broken for
co-annihilation.\footnote{Importantly, because co-annihilation
eliminates direct detection constraints, we encounter the interesting
scenario whereby a mono-jet signal may be seen in a region of
parameter space for which self-annihilating dark matter has already
been ruled out.}
It is thus relic density and collider searches that provide the most
relevant constraints on the co-annihilation scenario.

\begin{figure}
\centering
\includegraphics[width=0.3\textwidth]{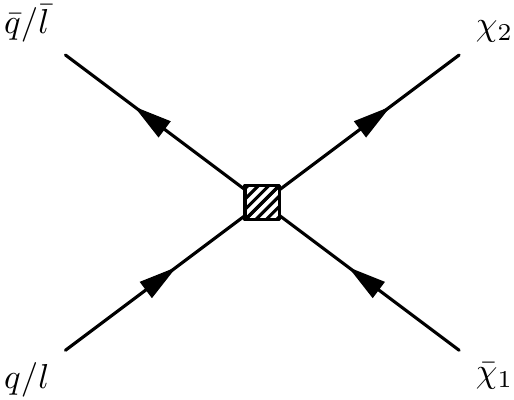}
\caption{Effective operators}
\label{fig:EffOp}
\end{figure}

\begin{figure}
\centering
\includegraphics[width=0.3\textwidth]{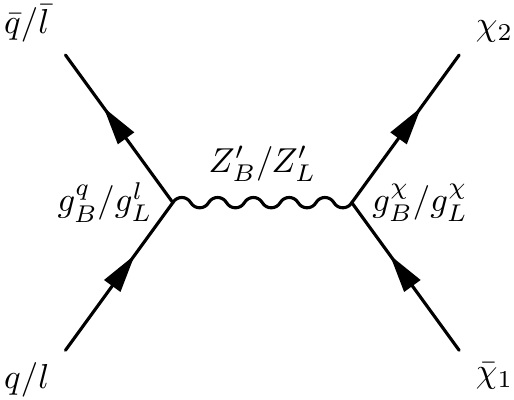}
\caption{UV Completion}
\label{fig:UVTh}
\end{figure}

However, as mentioned above, an EFT description may breakdown for high
energy collider experiments.  Therefore, we consider a simple UV completion
by introducing two neutral massive gauge bosons, $Z'_B$ and $Z'_L$,
associated with the spontaneous breaking of the gauged symmetries
$U(1)_B$ and $U(1)_L$ respectively.  We assume that spectator fermions
necessary to cancel the corresponding anomalies are massive enough to
evade current collider constraints.
The relevant interaction terms in the UV-theory are given by,
\begin{eqnarray}
\Delta \mathcal{L}_{UV}&=&g_B^q\bar{q}\gamma^{\mu}Z'_{B,\mu} q+g_L^l\bar{l}\gamma^{\mu}Z'_{L,\mu} l \nonumber \\
&+&g_B^{\chi}\left[\bar{\chi_2}\gamma^{\mu}Z'_{B,\mu} \chi_1+h.c+\bar{\chi_2}\gamma^{\mu}Z'_{B,\mu} \chi_2\right]\nonumber\\
&+&g_L^{\chi}\left[\bar{\chi_2}\gamma^{\mu}Z'_{L,\mu}\chi_1+h.c+\bar{\chi_2}\gamma^{\mu}Z'_{L,\mu}\chi_2\right],
\label{UVlagrangian} 
\end{eqnarray}
where $q$ and $l$ represent the SM quarks and leptons. The $Z'_B$
($Z'_L$) gauge boson couples to quarks (leptons) with strength $g_B^q$
($g_L^l$) and to dark sector particles with strength $g_B^\chi$
($g_L^\chi$).  Notice that in this case there are two effective
scales,
\begin{equation}
\Lambda_{12,B}=\Lambda_{22,B}=\frac{M_{Z'_{B}}}{\sqrt{g_{B}^q g_B^\chi}} \;, \quad 
\Lambda_{12,L}=\Lambda_{22,L}=\frac{M_{Z'_{L}}}{\sqrt{g_{L}^lg_L^\chi}} \; .
\label{eqn:LambdaDef}
\end{equation}

The $\bar{f}f \leftrightarrow \chi_i\overline{\chi_j}$ processes in the EFT and
UV complete descriptions are shown in Fig.~\ref{fig:EffOp} and
Fig.~\ref{fig:UVTh}, respectively.
For the parameters we choose, the EFT description will be approximately
valid for the relic density calculation as we describe in the next
section. However, for the LHC collider phenomenology, where the
momentum transfer can be large and the EFT breaks down, we use the
UV-complete theory.


\section{Dark matter relic density}
\label{sec:relic}

In the co-annihilation scenario we calculate the relic number density
at freezeout of $n=\sum_i n_{\chi_i}$.  The $\chi_2$ component will then
decay to the stable DM particle $\chi_1$, to form the relic DM in the
universe today.  The effective annihilation cross section of the $\chi_i$
to SM fermions, $f$, is~\cite{Griest:1990kh,Edsjo:1997bg}
\begin{eqnarray}
\sigma_{\rm eff} =\frac{1}{(g_{\rm eff}/4)^2}\big[ && \sigma_{\chi_1\bar{\chi}_1\to f\bar{f}} 
+ e^{-\Delta  x}(1+\Delta)^{3/2}\sigma_{\chi_1\bar{\chi}_2\to f\bar{f}} \nonumber \\
&&+ e^{-2\Delta  x}(1+\Delta)^{3}\sigma_{\chi_2\bar{\chi}_2\to f\bar{f}}\big]\; ,
\label{reliccross} 
\end{eqnarray}
where $x\equiv m_{\chi_1}/T$,
\begin{equation}
\Delta\equiv\frac{m_{\chi_2}-m_{\chi_{1}}}{m_{\chi_1}},
\end{equation}
and
\begin{equation}
g_{\rm eff}=4+4e^{-\Delta x}(1+\Delta)^{3/2}\;.
\end{equation}

We are interested in parameters for which the self-annihilation cross
section of $\chi_1$, $\sigma_{\chi_1\bar{\chi}_1}$is suppressed due to
the large UV cut-off $\Lambda_{11}$.  Unless the mass splitting
of $\chi_1$ and $\chi_2$ is extremely small, we expect the last term
in Eq.(\ref{reliccross}) to provide a negligible contribution to
$\sigma_{\rm eff}$ due to the double exponential suppression.  Working
under these assumptions the co-annihilation of $\chi_1$ and $\chi_2$
dominates the dark matter depletion in the early universe.  
We now determine the parameter values necessary to obtain the observed DM relic density.

In the usual thermal freeze-out scenario we can apply a
non-relativistic expansion of the annihilation cross section such that
the thermal average is given by $\langle \sigma_{\rm eff}
v\rangle\approx a_{\rm eff}+6b_{\rm eff}/x$.  The relic density is
then approximated to good accuracy by~\cite{Kolb:1990}
\begin{equation}
\Omega_{\chi_1} h^2=\frac{1.07 \times 10^9 {\rm GeV}^{-1} x_F}{g_*^{1/2}M_{Pl}(I_a+3\frac{I_b}{x_F})}\;,
\end{equation}
where $M_{Pl}=1.22\times 10^{19}$ GeV is the Planck mass, $g_*$ is
the total number of relativistic degrees of freedom at the freeze-out
temperature,
\begin{equation}
I_a=x_F\int_{x_F}^{\infty}dx \frac{a_{eff}}{x^{2}}\;,\qquad I_b=2x_{F}^2\int_{x_{F}}^{\infty}dx\frac{b_{eff}}{x^{3}}\;,
\end{equation}
and the freeze-out temperature is determined by,
\begin{equation}
x_{F}=\log\left(c(c+2)\sqrt{\frac{45}{8}}\frac{4}{2\pi^3}\frac{m_{\chi_1} M_{Pl}(a_{eff}+b_{eff}/x_F)}{\sqrt{g_{eff}x_F}}\right)\;,
\end{equation}
with $c\approx 1/2$. Thus it is clear that, in the EFT limit, the relic density $\Omega_{\chi_1}$ is completely determined by the parameters $m_{\chi_1}$, $\Delta$ and $\Lambda_{12}$.  
We implemented the model with $FeynRules$~\cite{Christensen:2008py} and generated model files for $MicroMEGAs$~\cite{Belanger:2006is} for the relic abundance calculation.
We then performed a scan over these parameters such that a relic density in approximate accordance with the Planck results was obtained, $\Omega_{DM}=0.1187\pm0.0017$~\cite{Ade:2013ktc}.

\begin{figure}
\includegraphics[width=1.0\columnwidth]{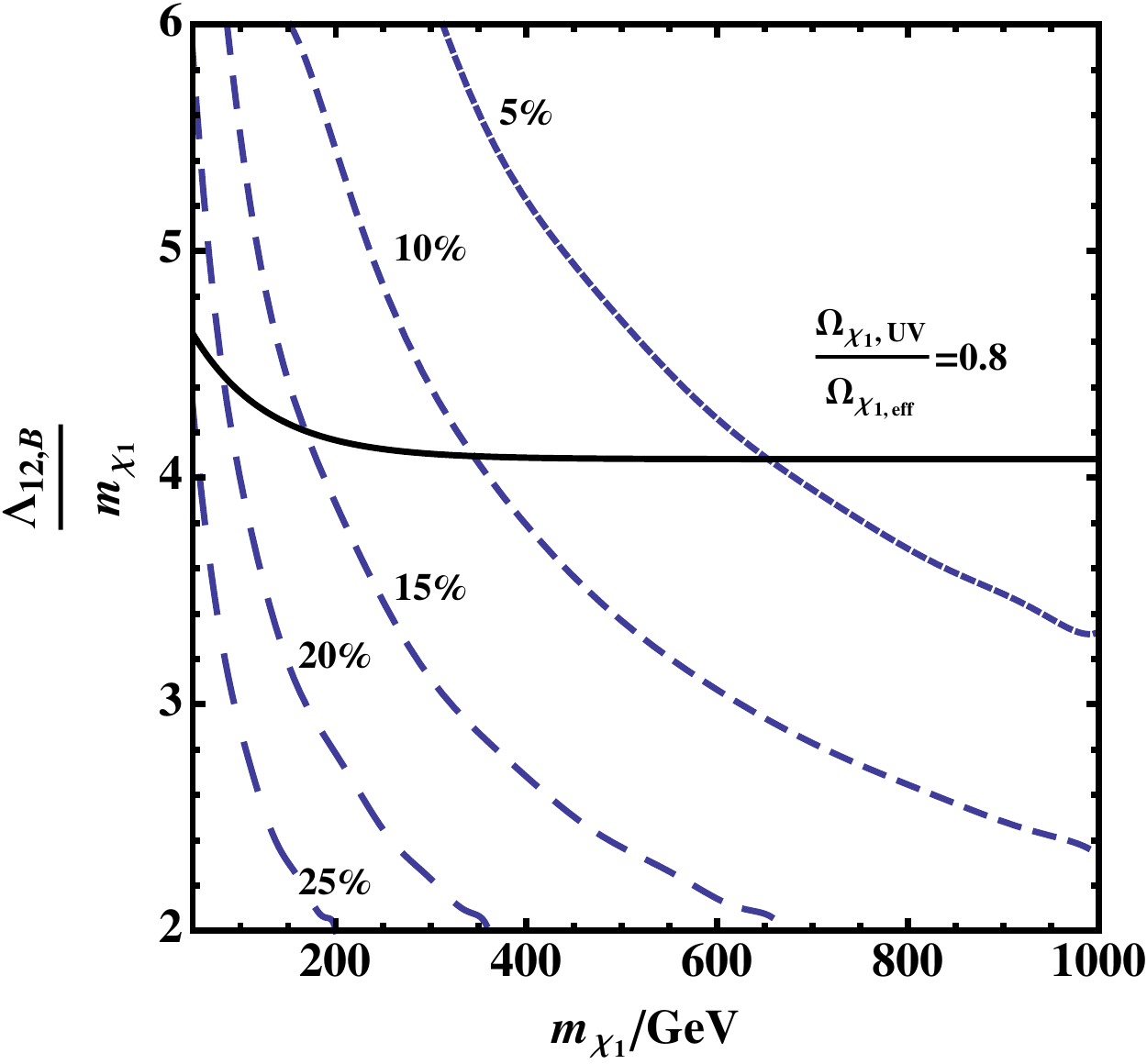}
\caption{Contours of the mass splitting $\Delta$ which satisfy the
  correct DM abundance in the EFT approximation.  Contours are shown
  with dashed lines for $\Delta=0.05$, 0.1, 0.15, 0.2 and 0.25, as
  indicated.  The solid line represents $\Omega_{\chi_1, {\rm UV}} /
  \Omega_{\chi_1,{\rm eff}} = 0.8$, where $\Omega_{\chi_1,{\rm UV}}$
  is calculated taking $g_B^q=g_B^\chi=1$. The EFT provides an
  adequate description for parameters above this line.}
\label{fig:relicplot}
\end{figure}

In Fig.~\ref{fig:relicplot} we show the results of the scan for dark matter mass between $50 \;\rm{GeV}$ 
and $1 \;\rm{TeV}$ and $\Lambda_{12,B}/m_{\chi_1}$ between 2 and 6, assuming that dark matter only couples to quarks, i.e. $\Lambda_L \gg \Lambda_B$.\footnote{When we turn to the collider phenomenology, we shall be interested in
dark matter which couples to both quarks and leptons.  Thus the relic
density calculations will be appropriately modified to take the
additional annihilation channels into account.}
The necessary mass splitting for successful co-annihilation is
typically $\Delta \lesssim 0.3$, and in Fig.~\ref{fig:relicplot} we
show contours corresponding to $\Delta= 0.05 - 0.25$.  For fixed
$\Lambda_{12,B} / m_{\chi_1}$, the co-annihilation cross section
decreases with increasing $m_{\chi_1}$, which in turn requires more
efficient co-annihilation, i.e. a smaller $\Delta$.  On the other hand,
for fixed $m_{\chi_1}$, the annihilation cross section also decreases
with increasing $\Lambda_{12}$, which again can be compensated with a
more efficient co-annihilation.

In Fig.~\ref{fig:relicplot} we also indicate the parameters for which
the EFT description is valid.  The parameters above the solid line satisfy
$\Omega_{\chi_1,{\rm UV}} / \Omega_{\chi_1,{\rm eff}} \geq$  0.8, where
$\Omega_{\chi_1,{\rm UV}}$ and $\Omega_{\chi_1,{\rm eff}}$ are the relic
densities calculated using the UV complete and EFT descriptions, respectively.  We see
that the EFT underestimates the relic density unless $\Lambda_{12}$ is
sufficiently large, $\Lambda_{12}/m_{\chi_1} \gtrsim 5$.\footnote{If $M_{Z'_{B,L}}\approx \sqrt{s}\approx 2m_{\chi_1}$ resonance effects become important which are not taken into account in the effective theory. }

In section~\ref{sec:collider} we shall consider collider signatures for
parameters consistent with the the DM relic density.
Note that because $\chi_1\overline\chi_2$ co-annihilation makes an
exponentially suppressed contribution to $\sigma_{\rm{eff}}$ (see
Eq.~(\ref{reliccross})), we require larger coupling constants than for
the standard case of a single dark sector species.  Correspondingly, the
collider production rates are larger.

\section{Couplings constraints from di-jet and di-lepton searches}
\label{sec:z'constraints}

As with many dark matter models, in addition to searches for dark
matter production, important constraints arise from direct production
of the particles which mediate the DM-SM interaction.  In our model
these mediators are the $Z'_L$ and $Z'_B$ particles.  We are thus led
to consider di-jet and di-lepton signals with no $\cancel{E}_T$, mediated by $Z'$
exchange.

In an EFT description the relevant
four-fermion effective operators are
\begin{equation}
\frac{1}{\Lambda_{l}^2}\bar{l}\gamma^{\mu}l\bar{l}\gamma_{\mu} l\;, \qquad \frac{1}{\Lambda_{q}^2}\bar{q}\gamma^{\mu} q \bar{q} \gamma_{\mu} q \;,
\label{Z'op}
\end{equation} 
where $\Lambda_l=M_{Z'_{L}}/|g_Z^l|$ and
$\Lambda_{q}=M_{Z'_{B}}/|g_B^q|$. There have been several experiments that
constrain the allowed parameter space of operators of the form of
Eq.~(\ref{Z'op}).  
For example, $\Lambda_l$ is constrained by dilepton searches at LEP
II.  For $m_{Z'_L} > {\rm 200\; GeV}$, dilepton production through an
off-shell mediator results in the bound $g_L^l\lesssim
g_{L,max}^l\equiv 0.044\times (m_{Z'_{L}}/200\; {\rm
  GeV})$~\cite{LEP:2003aa,Buckley:2011vc}.

On the other hand, the $Z'_{B}$ parameters are constrained by di-jet
searches at hadron colliders such as UA2 at the CERN SPS collider, CDF
at the Tevatron, and ATLAS and CMS at the LHC.  For example,
Ref.~\cite{Dobrescu:2013cmh} reports limits on the $g_{Z'_{B}}$ -
$M_{Z'_{B}}$ coupling-mass plane, arising from di-jet resonance
searches. (Note that di-jet resonances, corresponding to on-shell
$Z'_B$ production, cannot be described in terms of an EFT.)  However,
these limits assume the new $Z'_{B}$ boson decays only to SM
particles.  If the $Z'_{B}$ boson can also decay to dark sector
particles, the limits from the di-jet resonances searches are weakened
because the on-shell $Z'$ can now decay to invisible final states.  If
the $Z'_{B}$-width is narrow, the s-channel production factorizes from
the decay, $\sigma(jj) = \sigma_{Z'}\times{\rm Br} (Z'\rightarrow jj)$.
Although the di-jet branching ratio will be reduced by the presence of
other decay channels, we can safely assume that the di-jet acceptance
is not altered.  Therefore, we see
that the effect of the additional $Z'$ decay modes is to increase the
maximum allowed coupling to quarks by a factor of $1/\sqrt{ \rm{Br}
}$, namely
\begin{eqnarray}
g_{B,max}^q  && = g_B^q\frac{1}{\sqrt{Br(Z'_{B}\to jj)}} 
\;, \\
&& 
=  g_B^q \left[ 1 + \frac{\Gamma(Z'_{B}\to \chi\overline{\chi})}{\Gamma(Z'_{B}\to jj)}\right]^{1/2}
\;, \nonumber \\
&& \approx  g_B^q\left[1+\frac{r_{B}^2}{N_c N_f}\left(1+2\frac{m_{\chi_1}^2}{M_{Z'_{B}}^2}\right)\sqrt{1-\frac{4m_{\chi_1}^2}{M_{Z'_{B}}^2}}\right]^{1/2} \;, \nonumber
\end{eqnarray}  
where $r_{B}=g_B^\chi/g_B^q$ is the coupling ratio, $N_{c}$ and
$N_{f}$ are the number of color and flavour states the $Z'_{B}$ can
decay into, and we have neglected SM-fermion masses since $m_{f,SM}\ll
m_{\chi_1},M_{Z'_{B}}$.

We shall choose example parameters that are consistent with the pure di-jet and di-lepton constraints, and which also produce the correct DM relic density.  For this purpose, it is useful to write the couplings $g_B^q$ and $g_L^l$ as functions of the coupling ratios $r_B$ and $r_L$ respectively,
\begin{equation}
g_{(B,L)}^{(q,l)}=\frac{M_{Z'_{B,L}}}{\Lambda_{12,(B,L)}r_{B,L}^{1/2}}\label{gzp}.
\end{equation} 
In order to maximise the possible signals, we choose couplings that
are close to, but do not saturate, the current bounds.  Specifically,
we take $g_{L,max}^l-g_{L}^l=0.01$ and
$g_{B,max}^q-g_B^q=0.1$.  We will divide our analyses in two groups.
In the first group, we only allow $U(1)_L$ and $U(1)_B$ couplings of the $Z'$s to SM particles and
hidden sector particles that are below 1. We refer to this group as the {\it weak coupling}
group. The second group consists of points in parameter space where the  $U(1)_L$ and $U(1)_B$
$Z'$ couplings to hidden sector particles, $g^{\chi}_{B}$ and $g^{\chi}_{L}$,  are allowed to be larger than 1. 
We refer to this group as the {\it strong coupling} group. For the weak coupling group
we fixed the $Z'_{L}$ mass to be $M_{Z'_L}=250$ GeV, which allows for smaller $g^{\chi}_{L}$-couplings,
while for the strong coupling group we choose $M_{Z'_L}=550$ GeV. 
Using Eq.~(\ref{gzp}), we can then obtain masses and couplings
that satisfy both the relic density and collider constraints.  Some
representative parameters are shown in Table~\ref{tab:num1}.

\renewcommand{\arraystretch}{1.2} 
\begin{table*}[tb]
\begin{center}
    \begin{tabular}{|c||c|c|c|c|c|c|c|c|c|c|c|c|c|}
\hline
& $m_{\chi_1}$  & $m_{\chi_2}$  & $m_{Z'_B}$  & $g_B^q$  & $g_B^\chi$ & $m_{Z'_L}$  &  $g_L^l$ & $g_L^\chi$ & $\Lambda_{l}$  &
$p_{T}(l_1)$   & $M(l^{+} l^{-})$   & $\sigma_{p p\to \bar{\chi}_i \chi_j}$ &   $\sigma_{p p\to \bar{\chi}_1 \chi_1 l^{+}l^{-}}$\\
example & (GeV) & (GeV) & (GeV) &  & & (GeV) &  &  & (GeV) & (GeV) & (GeV) &   14 TeV (fb) &   14 TeV (fb) \\
\hline\hline
1. & 250 & 270 & 525 & 0.15 & 0.80 & 250 & 0.045 & 0.66 &  1450  & $<30$  & $<$ 20 &  $6597$ & $552$ \\
\hline
2. & 300 & 321 & 625 & 0.14 & 0.89 & 250 & 0.045 & 0.53 & 1620   & $<30$ & $<$ 20  & $3694$  & $376$ \\
\hline
3. & 400 & 420 & 825 & 0.18 & 0.68 & 250 & 0.045 & 0.32 & 2080  & $<30$  & $<$ 20 & $905$  &  $102$ \\
\hline
4. & 600 & 612 & 1700 & 0.23 & 0.98 & 250 & 0.045 & 0.15 & 3000 & $<20$  & $<$ 15 & $442$ & $52$ \\
\hline\hline
5. & 400 & 432 & 1375 & 0.21 & 2.2 & 550 & 0.11 & 0.8  & 1840  & $<60$  & $<$ 30 &   $2285$ &   $186 $ \\
\hline
6. & 500 & 530 & 1500 & 0.18 & 1.83 & 550 & 0.11 & 0.52 & 2300  & $<60$  & $<$ 30  &     $1103  $ &     $104$ \\
\hline
7. & 600 & 630 & 1475 & 0.16 & 1.61 & 550 & 0.11 & 0.36 & 2760  & $<40$   & $<$ 30  &   $852$ &   $70$ \\
\hline
8. & 700 & 728 & 1425 & 0.12 & 1.51 & 550 & 0.11 & 0.26 & 3220 & $<30$    & $<$30 &  $193$ & $16$\\ 
\hline
  \end{tabular} \end{center} \caption{Coupling parameters, cross sections, and kinematic cuts for eight
   sets of example parameters.  The second and the third column are
   the masses of the two co-annihilating particles, while the fourth
   to the ninth columns show the $Z_B'$ and $Z_L'$ mass and coupling
   parameters.
We also show the cuts on the leading lepton-$p_T$, the invariant mass
   of the dileptons and the $\Delta R$ separation between the leading
   lepton and the leading jet.
 The last column is the cross
   section for $pp\rightarrow \chi_i\bar{\chi}_j + j\rightarrow \chi_1\bar{\chi}_1 l^+ l^- + j $ at the 14 TeV LHC.
   \label{tab:num1}}
\end{table*}


\section{Collider Phenomenology}
\label{sec:collider}

\subsection{Signal processes}

We now examine the new LHC signals that arise from
$\overline\chi_1\chi_2$ production followed by the decay
$\chi_2 \rightarrow \chi_1 + SM$.
The lowest order $\overline\chi_1\chi_2$ production process at the LHC
is
\begin{equation}
pp \rightarrow  \overline\chi_1 \chi_2 \rightarrow \overline\chi_1\chi_1 + SM.
\label{eq:chichiff}
\end{equation}
In this case the only visible particles in the final state are the SM
states ($\overline{q}q$ or $\overline{l}l$) produced through the decay of the $\chi_2$.
The signal for this process is thus di-leptons or di-jets plus
$\cancel{E}_{T}$.
However, given that successful co-annihilation
requires a relatively small mass difference between $\chi_2$ and
$\chi_1$, $\Delta m \lesssim 0.3 \ m_{\chi_1}$, the SM particles have
relatively soft energies and the remnant hidden particles $\chi_1$ and
$\overline\chi_1$ are approximately back-to-back, leading to a small
net $\cancel{E}_{T}$.
This is crucial for the possible detectability of this signal. At a
hadron collider the irreducible $Z+jets$ background, with soft jets
from the underlying QCD processes, provides an enormous number of
events with dileptons or dijets and small $\cancel{E}_{T}$, even when
demanding a $Z$-veto by rejecting events whose dilepton or
dijet invariant mass is close to the $Z$ boson pole mass.  The process in
Eq.(\ref{eq:chichiff}) would be hidden by this large background at the
LHC.~\footnote{In a lepton collider such as the ILC it is conceivable
that the dijet or dilepton plus $\cancel{E}_{T}$ signal may be
discoverable due to the absence of underlying QCD process, making it a
cleaner environment.}.

\begin{figure}
\includegraphics[width=0.35\textwidth]{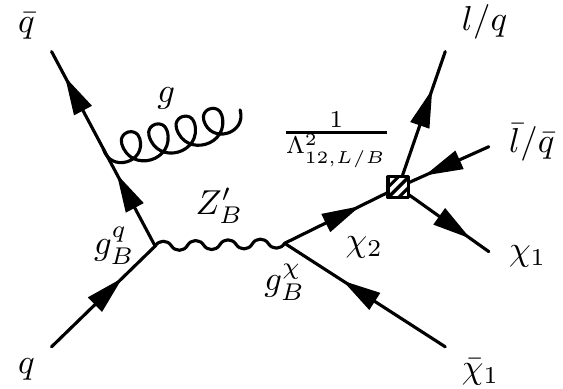}
\caption{Contribution to the jet plus $\cancel{E}_T$ plus dilepton (or diquark) signal at the LHC.  Additional diagrams with other initial state partons are not shown.}
\label{fig:signal}
\end{figure}

In order to have any chance of observing $\overline\chi_1\chi_2$
production in a hadron collider, it becomes imperative to look for
processes that breaks the back-to-back alignment between the hidden
particles and therefore lead to a substantial $\cancel{E}_{T}$. This
alignment rupture is in fact produced by a hard jet in the form of
initial state radiation (ISR), as is the case for monojet searches.
In this case the hidden particles recoil against the hard jet and a
large amount of $\cancel{E}_{T}$ can be produced.  Therefore, the
process of interest at the LHC is
\begin{equation}
pp \rightarrow  \overline{\chi_1}\chi_2 + j \rightarrow \overline\chi_1\chi_1  + SM + j,
\end{equation}
where the ISR jet is hard.
Given that we have two mediators, the possible $\chi_2$ decay chains are
\begin{eqnarray}
&& \chi_2 \to \chi_1 + {Z'_{B}}^* \to \chi_1 + q +\bar{q} \;, \\
&& \chi_2 \to \chi_1 + {Z'_{L}}^* \to \chi_1 + l +\bar{l}  \;,
\end{eqnarray} 
and hence the final states will be a hard jet from ISR in addition to
$\cancel{E}_{T}$ and either di-jets or di-leptons.  A representative
diagram is shown in Fig.~\ref{fig:signal}.
Again, due to the small mass splitting $\Delta m$, the energy of the
di-jets or di-leptons will be relatively low.~\footnote{Note that for
the decay $\chi_2\to\chi_1+SM$, the small $\chi_2$ - $\chi_1$ mass
difference implies that the $Z'$ propagator has a low momentum
transfer and thus the EFT description will be valid.}.  As it would be
challenging to identify di-jet signals with low $p_T$, due to the
large QCD backgrounds encountered at the LHC, we shall concentrate on
the di-lepton channel.

The production cross sections of $\overline{\chi_1}\chi_2$ plus
$\overline{\chi_2}\chi_1$ are shown in Table~\ref{tab:num1}.  Also
shown are the cross sections for $\overline{\chi_1}\chi_1 l^+l^-$, which
are related to the previous cross sections by the $\chi_2$ branching
ratio to leptons, which is typically $O$(10\%) for our parameters.

\renewcommand{\arraystretch}{1.2} 
\begin{table*}[t]
  \begin{center}
       \begin{tabular}{|c|c|c|c|c|}
 
      \hline
       Cuts &      Signal (S) &   Background (B) &        Significance ($S/\sqrt{S+\Delta B}$)\\
      \hline
       \begin{tabular}{c}
       $p_{T}(l)>10$ GeV , $|\eta_{lep}|<2.5$, \\
        $\Delta R_{l^{+}l^{-}}>0.4$ , $\Delta R_{lj}>0.4$
         \\ $M(l^{+} l^{-})> 5$ GeV
        \end{tabular}
        &      7520  &  1062935  &         0.10 \\
       \hline
          $p_T(j_1)> 150$ GeV &         1650  &     428354  &        0.04  \\       
       \hline
            $\cancel{E}_T>120$ GeV &       1079  &   22090  &         0.61 \\      
      \hline
           $M(l^{+} l^{-})<20$ GeV &        55  &       85  &         3.8 \\
      \hline
          $N(b) =0$ & 53 & 38 & 5.2 \\
      \hline
       $p_T(l_1)<30$ &      52 &        14  &        6.3 \\
\hline
    \end{tabular}
     \end{center}
     \caption{The cut-flow chart for example \#2 with $\mathcal{L}=20$ fb$^{-1}$.  The cuts are sequential. 
\label{tab:num2}}
 \end{table*}

\renewcommand{\arraystretch}{1.2} 
\begin{table*}[htb]
  \begin{center}
   \caption{ Signal and Background after cuts for the example parameters.\label{tab:num3}}
       \begin{tabular}{|c|c|c|c|c|c|}
      \hline
       Example \# &  $\mathcal{L}$ at $s=14\; \rm{TeV}$ &      Signal (S) &   Background (B) &       $S/B$ &  $S/\sqrt{S+\Delta B}$\\
      \hline\hline
            1. &   $\mathcal{L}=20$ fb$^{-1}$ &       67 &    14      & 4.8 &      7.4 \\
      \hline
             2. & $\mathcal{L}=20$ fb$^{-1}$ &        52  &       14      &   3.7 & 6.3 \\
      \hline
           3. &  $\mathcal{L}=200$ fb$^{-1}$ &        106   &     137   &       0.77 &  5.1\\
      \hline
           4. &  $\mathcal{L}=300$ fb$^{-1}$ &        23   &     41    &       0.56  & 2.6\\
      \hline
          5. &  $\mathcal{L}=20$ fb$^{-1}$ &        159  &       93   &    1.7 & 8.6 \\
       \hline
          6. &  $\mathcal{L}=20$ fb$^{-1}$ &        81   &       93   &    0.87  & 5.0\\
       \hline
          7. &  $\mathcal{L}=20$ fb$^{-1}$ &        51   &       63   &    0.80 & 4.1 \\
       \hline
          8. &  $\mathcal{L}=300$ fb$^{-1}$ &        114   &       538   &    0.21 & 1.9 \\
    \hline
    \end{tabular}
\end{center}
\end{table*}

\subsection{Backgrounds and event selection at the LHC}

We now perform a detailed calculation of the signal and background for
the process $pp \to \overline\chi_1\chi_2 j \to \overline\chi_1 \chi_1
j l^+l^-$, which is observed as a hard jet, large $\cancel{E}_{T}$, and two
same flavour opposite-sign leptons ($l^{+}l^{-}$), where by leptons we
mean light leptons $l=\{e,\mu\}$.

We use $MadGraph\;5$~\cite{Alwall:2011uj} to simulate signal and all background
events. The built-in $Pythia$~\cite{Sjostrand:2007gs} and $Delphes$~\cite{deFavereau:2013fsa} in $MadGraph\;5$ are used
to simulate the hadronization, showering and detector effects.
We adopt the MLM jet matching algorithm and set  $xqcut=15\ \rm{GeV}$.
We perform all analysis with
$MadAnalysis\;5$~\cite{Conte:2012fm}.  Our intention is not to
optimize the analysis to set precise constraints, but to instead
illustrate the potential for these dark matter models to be
constrained by the LHC experiments, and indicate which regions of
parameter space can be probed with forthcoming LHC data.

\begin{figure*}[ht]
\centerline{ \hspace*{-0.5cm}
\includegraphics[width=0.9\columnwidth]{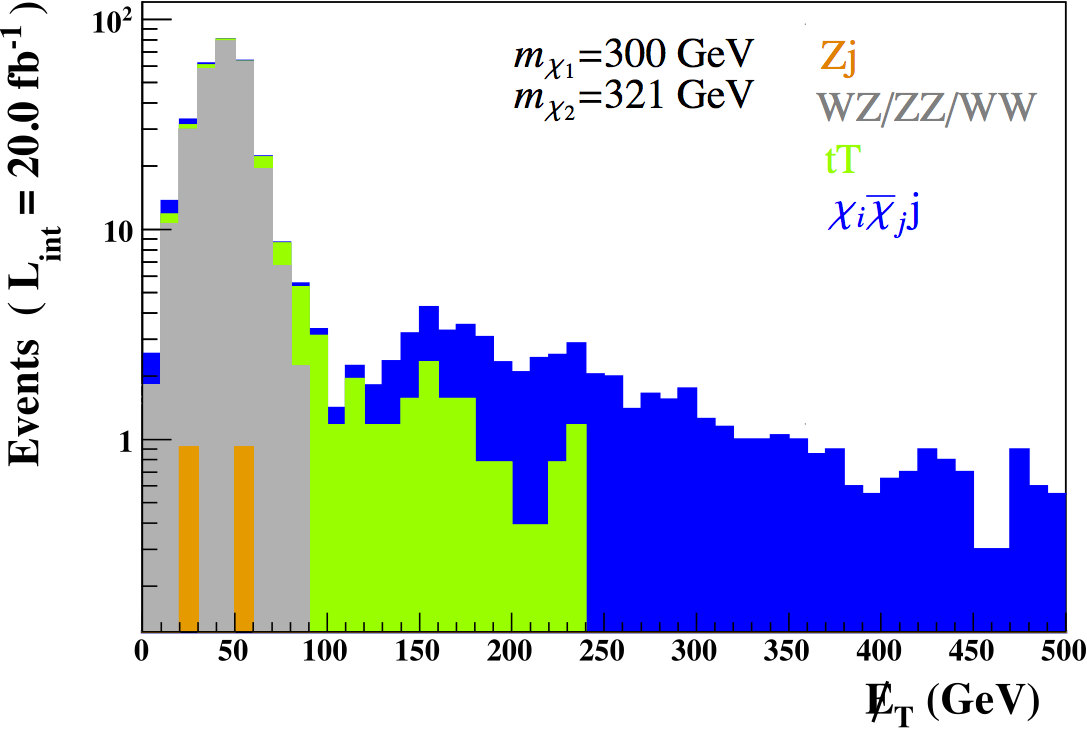}
\hspace*{0.5cm}
\includegraphics[width=0.9\columnwidth]{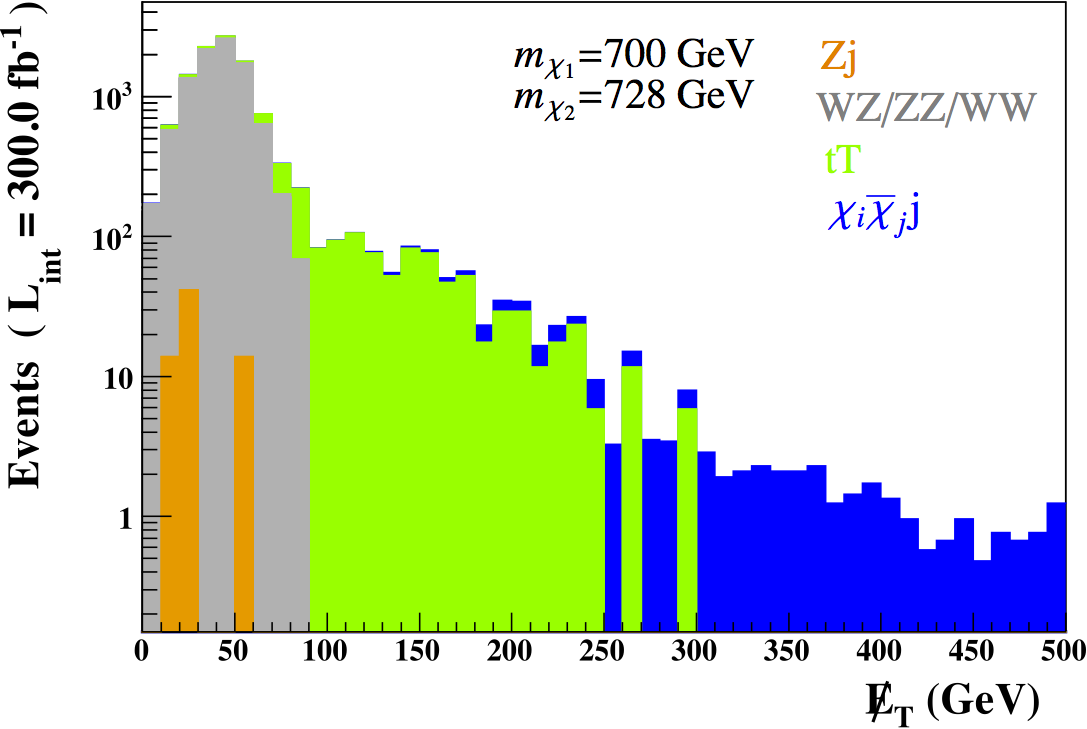}
}
\caption{ Histograms of the $\cancel{E}_T$ for example \#2 (left) and \#8(right) with all the cuts applied except the one 
on $\cancel{E}_T$, $\cancel{E}_T>120$GeV.}

\label{fig:met}
\end{figure*}

\begin{figure*}[ht]
\centerline{ \hspace*{-0.5cm}
\includegraphics[width=0.9\columnwidth]{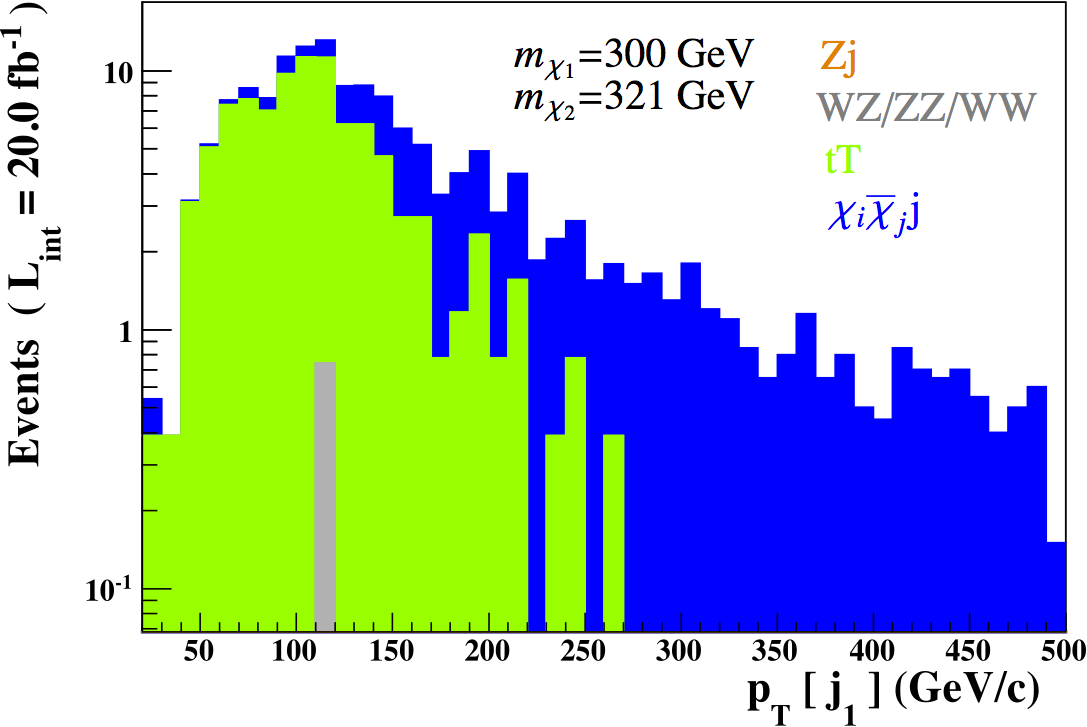}
\hspace*{0.5cm}
\includegraphics[width=0.9\columnwidth]{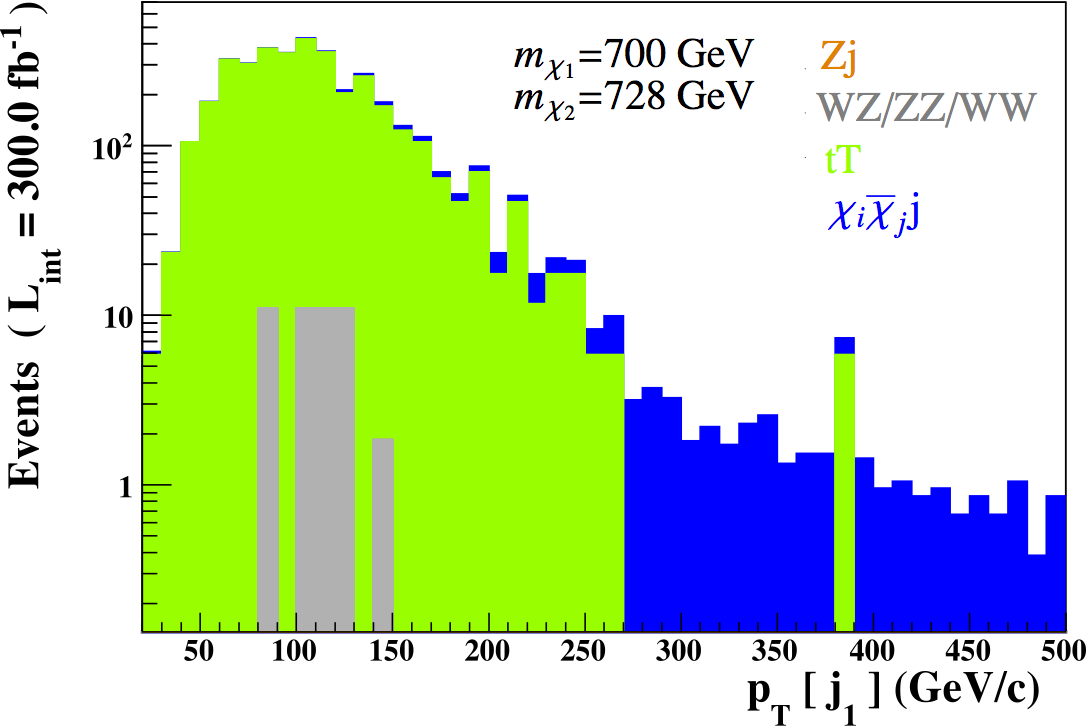}
}
\caption{ Histograms of the leading jet-$p_T$  for example \#2 (left) and \#8(right) with all the cuts applied except the one 
on the leading jet-$p_T$, $p_T(j_1) > 150$ GeV.}
\label{fig:ptj1}
\end{figure*}

\begin{figure*}[ht]
\centerline{ \hspace*{-0.5cm}
\includegraphics[width=0.9\columnwidth]{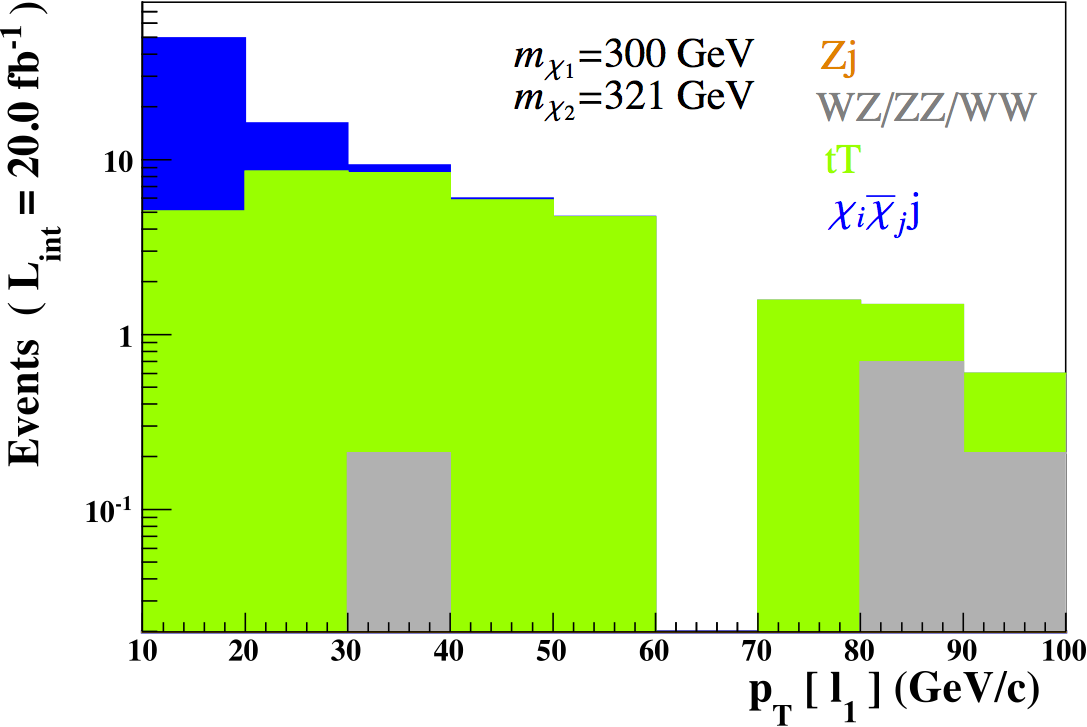}
\hspace*{0.5cm}
\includegraphics[width=0.9\columnwidth]{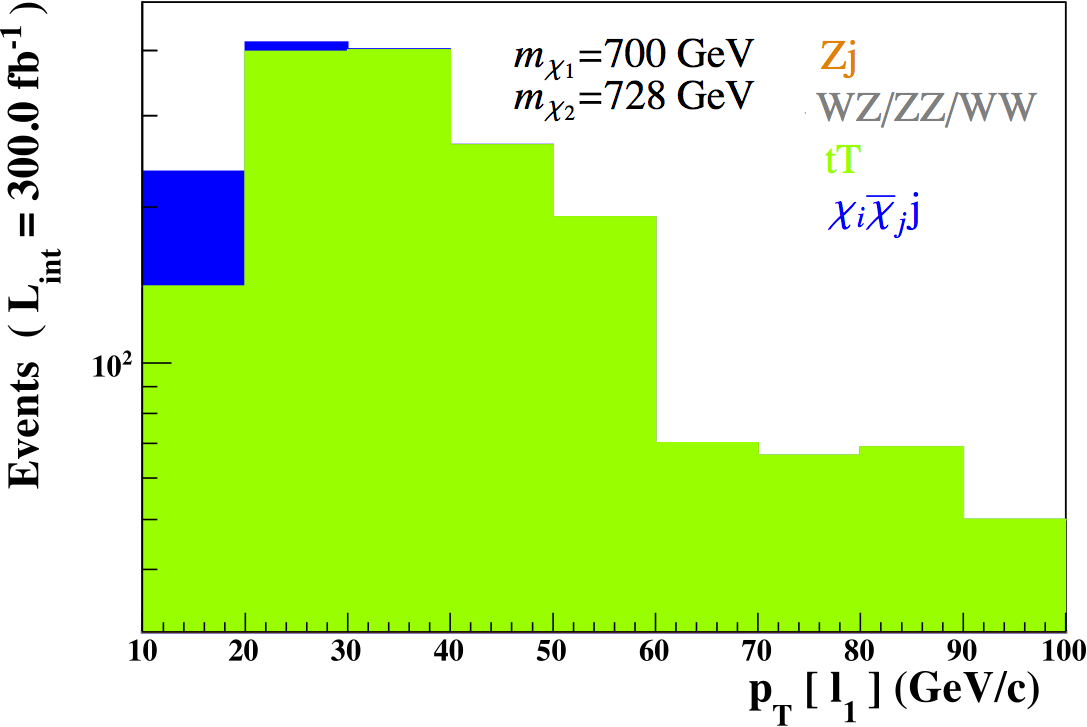}}
\centerline{ \hspace*{-0.5cm}
\includegraphics[width=0.9\columnwidth]{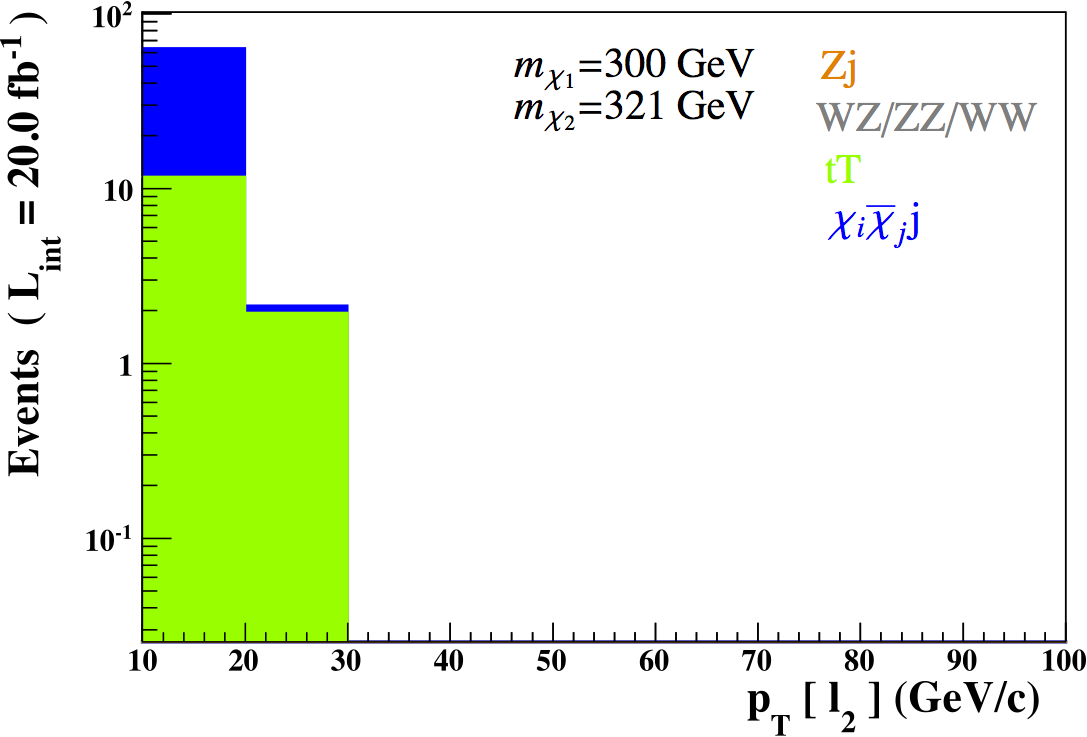}
\hspace*{0.5cm}
\includegraphics[width=0.9\columnwidth]{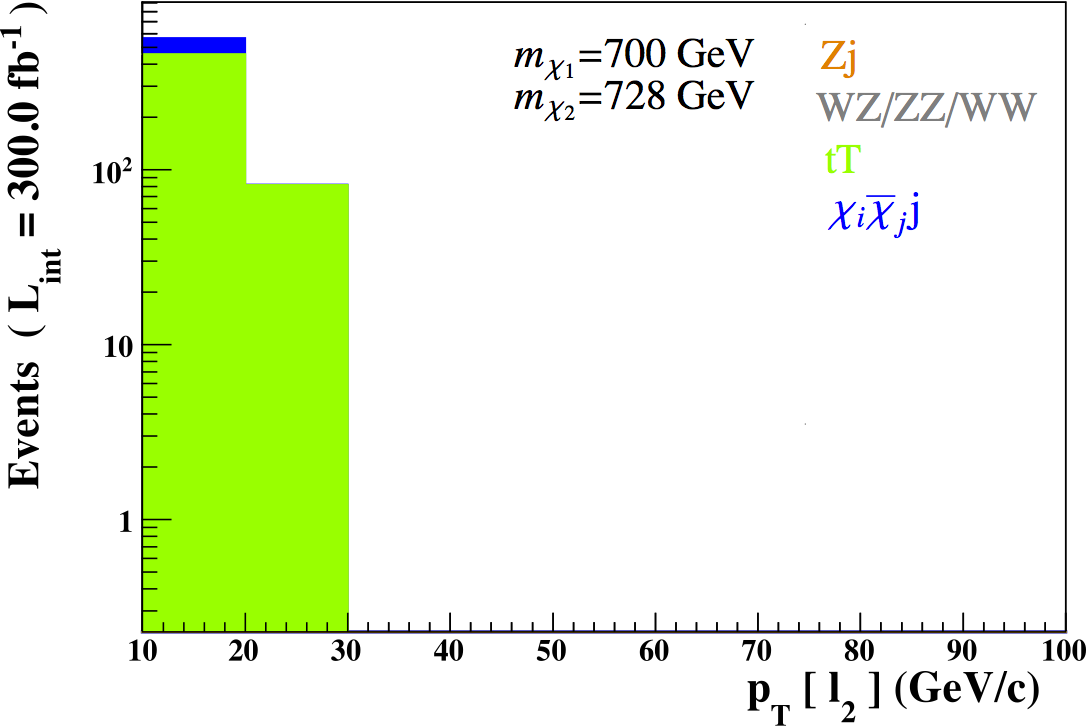}
}
\caption{ 
Top: histograms of the leading lepton-$p_T$ for example \#2 (left) and \#8 (right) with all cuts applied 
except the ones on the leading lepton-$p_T$, $ 10 \rm{GeV} < p_T(l_1) < 30 \rm{GeV}$ for both \#2 and \#8.
Bottom:  histograms of the second leading lepton-$p_T$ for example \#2 (left) and \#8 (right)  with all the cuts applied.}
\label{fig:ptl}
\end{figure*}

\begin{figure*}[ht]
\centerline{ \hspace*{-0.5cm}
\includegraphics[width=0.9\columnwidth]{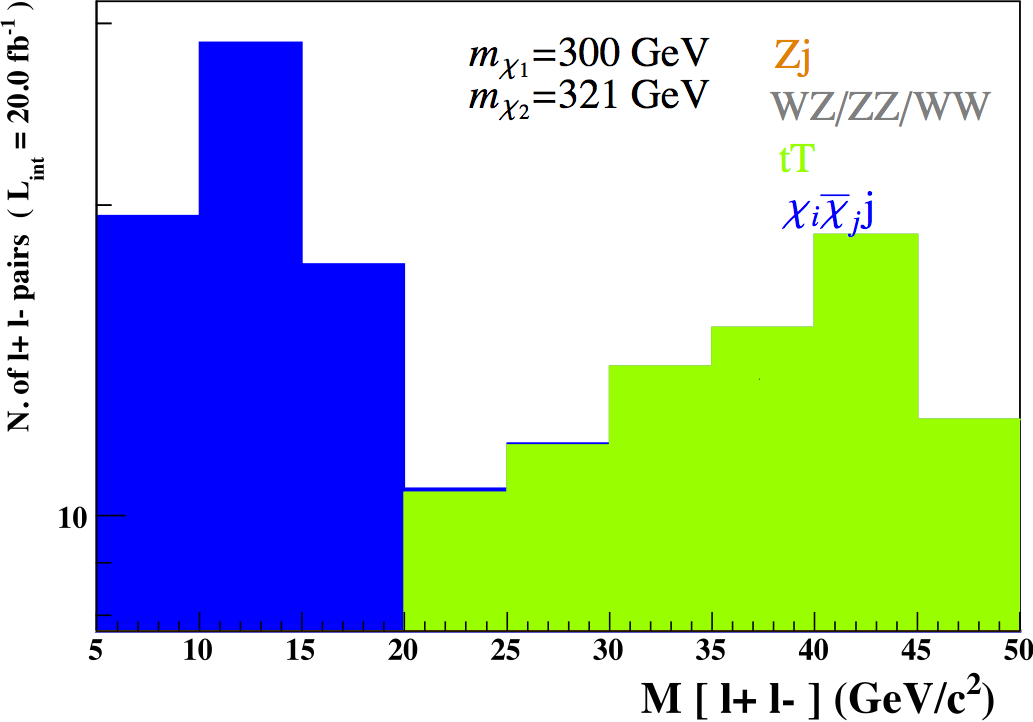}
\hspace*{0.5cm}
\includegraphics[width=0.9\columnwidth]{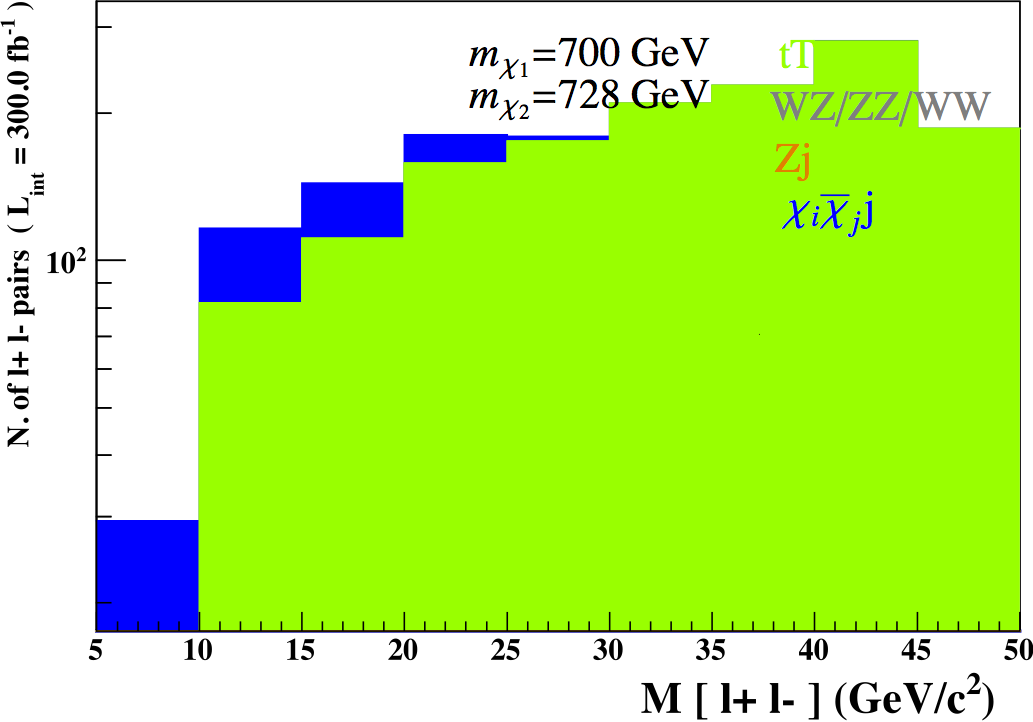}
}
\caption{ Histograms of the invariant dilepton mass for example \#2 (left) and \#8 (right)
with all cuts applied except the one on the invariant dilepton mass, $5 \rm{GeV} < M(l+,l-) <  20\rm{GeV}$  for \#2 (left)
and $5 \rm{GeV} < M(l+, l-) < 30 \rm{GeV}$ for \#8 (right).}
\label{fig:invdilep}
\end{figure*}

As mentioned earlier, to overcome the large background from
$Z(\to l^{+}l^{-})+jets$ we demand a hard jet and large
$\cancel{E}_T$, which makes this background
negligible~\cite{TheATLAScollaboration:2013oia}. Specifically, we
choose $p_T(j)>150$ GeV, and $\cancel{E}_T > 120$ GeV. We verified this in our analysis
as can be seen in Figs.~\ref{fig:met}--\ref{fig:invdilep} where one can see that the combination of the requirements
of a hard jet and large $\cancel{E}_T$ make this background negligible. Furthermore, we impose a minimum cut on the invariant mass of the dileptons $M(l^+
l^-)>5 $ GeV in order to veto background events from $J/\psi$ decays.

Another important background is the top pair production which subsequently decays via
 $t\bar{t}\to b\bar{b}l^+l^-\nu_l\bar{\nu}_l$. We can eliminate most of the $t\bar{t}$ background by demanding
 the leading lepton have $p_T<30-60 \rm{GeV}$ and that the $b$-jet multiplicity be zero. 

The main SM backgrounds remaining are  diboson pair
production,  $ZZ\to
l^+l^-\bar{\nu}_l \nu_l/l^+l^- j j$,  $WW \to l^+l^-\bar{\nu}_l \nu_l$ and $W^\pm Z\to l^+l^-l^\pm \nu$ , where
the leptons are of the same flavour and in the last process one of the leptons is missed. The hard jet can either come from the process itself or the underlying events. Since the diboson backgrounds are all electroweak processes, they
have much smaller cross section compared with the $t\bar{t}$ and $Z+jets$ background and
are thus subdominant.

%

We trigger our events by demanding $\cancel{E}_T>120$ GeV and
$p_T(j)>150$ GeV.  We adopt conservative requirements for the
charged leptons, $p_{T}(l)>10$ GeV, $|\eta|<2.5$, $\Delta
R(l^{+},l^{-})>0.4$ and $\Delta R(l,j) > 0.4$, where $\Delta R$ is the separation in the
$\eta$-$\phi$ plane such that the leptons are isolated~\footnote{In principle, because we trigger on the
$\cancel{E}_T>$ and the jet $p_T$, we could include lower momentum
leptons, $p_T(l)\gtrsim 6$ GeV, which would further boost our
signal.}.  We also demand the jet to have $|\eta|<5$. 
The remaining cuts were chosen to suppress as much background as
possible without diminishing the signal.  The kinematic observables
upon which we impose cuts are: the invariant mass of the leptons
$M(l^{+}l^{-})$ and $p_T$ of the leading lepton. We also veto
the events with nonzero $b$-jet multiplicity.
%
When considering high luminosities, in order to further reduce the
background, we also imposed an upper bound on the leading lepton,
$p_{T}(l_1)$.
We list in Table \ref{tab:num1} the corresponding cuts for the LHC at
14 TeV, for the sample DM parameters considered.  
In Table~\ref{tab:num2} we show, for the \#2 example parameters, how
each cut affects signal and background at 14 TeV and a luminosity of
$\mathcal{L}=20$ $\rm{fb}^{-1}$. 

%

%
%

%
%


%
%

%

%
%

In Figs.~\ref{fig:met}, \ref{fig:ptj1}, \ref{fig:ptl} and \ref{fig:invdilep} we show histograms of the kinematic variables
after all cuts except the one on the kinematic variable itself, for the example \# 2 at $\mathcal{L}=20$ fb$^{-1}$ 
and example \# 8 at $\mathcal{L}=300$ fb$^-1$. 
 Note from Fig.~\ref{fig:met} that
despite demanding a large $\cancel{E}_T$, there are still a sizable
number of events due to the balance between the ISR jet $p_T$ (which
tends to be hard) and the $p_T$ of the $\chi$.  Due to the small mass
difference $\Delta m \ll m_{\chi_1} \approx m_{\chi_2}$, the lepton
momentum is typically soft (see Fig.~\ref{fig:ptl}).  For the same
reason the invariant mass of the dileptons tends to be comparable to
$\Delta m$, as seen in Fig.~\ref{fig:invdilep}.
%

We define our signal significance via the ratio $S/\sqrt{S+\Delta B}$ with
\begin{equation}
\Delta B =\sqrt{\sum_i \left[ B_i +\left(\beta_i B_i\right)^2\right]},
\end{equation}
where $\beta_i$ is set to be $10\%$ to account for the systematic error.
 In Table~\ref{tab:num3} we list the signal (S), background (B), the ratio of signal over background and
the significances for different example parameters, considering the
minimal luminosity ($\mathcal{L}\lesssim 300$ fb$^{-1}$) needed to
obtain a significance of order 5 (or the largest possible for
$\mathcal{L}= 300$ fb$^{-1}$).
%

Based on the sensitivities shown in Table~\ref{tab:num3}, we conclude that the 14 TeV LHC does
have the capability to identify a significant signal.
For the {\it weak coupling} group (examples \#1-4), dark matter masses
up to $\sim$ 400 GeV would be probed in the first 200 fb$^{-1}$ of
luminosity at the 14 TeV LHC, while it may be possible to get a strong hint for
a dark matter mass of 600 GeV with  $\mathcal{L}=300$ fb$^{-1}$.
For the {\it strong coupling} group
(examples \#5-8), dark matter masses $m_{\chi}\lesssim$ 700 GeV would be probed
in the first 300 fb$^{-1}$ of luminosity at the 14 TeV LHC.
For masses $m_{\chi_1}\gtrsim 750$ GeV, the production cross-sections become too
small and therefore larger luminosities are needed. It is, however,
important to stress that scenarios with dark matter masses below
$\sim$ 250 GeV are already constrained by LHC measurements and even
with 20 fb$^{-1}$ of luminosity we should be able to detect (or rule
out) scenarios with dark matter masses below 300 GeV.

Note that all our example parameters satisfy the relic density
constraint, in the region where the EFT is valid. For sufficiently
large masses ( $\agt$ 1 TeV) and reasonable choices for the coupling
constants,
the correct
relic density cannot be explained unless we work in a region of
parameter space for which the EFT is not valid, for instance, close to
the resonance at $2 m_\chi = M_Z'$. Of course, if we relax the relic
density requirements, a wider range of collider cross sections are
possible.

\subsection{Other signals}

The co-annihilation model will also lead to mono-jet plus $\cancel{E}_T$
signals.  These arise from the invisible decays of the $\chi_2$ to
neutrinos, or decays to quarks or leptons which are too soft to be
identified.
Both search strategies (mono-jets plus $\cancel{E}_T$ and monojets plus dileptons
plus $\cancel{E}_T$) provide complementary information in the hunt for DM at the
LHC.
The latest CMS mono-jet analysis~\cite{CMS-PAS-EXO-12-048} places the
constraint $\Lambda_{12,B}\gtrsim 900$ GeV for the DM masses we
consider in Table~\ref{tab:num1}, and all our example parameters
satisfy this bound.  However, to properly determine the mono-jet
limits on our model, the process should be simulated using the
UV-complete theory.  Note, however, that
Ref.~\cite{An:2012va,An:2012ue} compared monojet plus $\cancel{E}_T$ constraints,
with di-jet constraints of the type discussed in
section~\ref{sec:z'constraints}, for models in which DM interactions
are mediated by light Z' bosons.  The di-jets analyses were found to
usually provide the more stringent constraints.

Finally, the possibility of pair production of $\overline\chi_2\chi_2$
presents further interesting signals.  The process
$pp \to \overline\chi_2\chi_2 \to l^+l^-l'^+l'^-$ results in two
pairs of opposite sign-leptons, plus $\cancel{E}_{T}$.
If we assume equal $\chi_1-\chi_2$ and $\chi_2-\chi_2$ couplings, as in
Eq.~(\ref{UVlagrangian}), the cross section for $\overline\chi_2\chi_2$
production will be comparable to that for $\overline\chi_1\chi_2$.
From the relic density point of view, the $\overline\chi_2\chi_2$
annihilation channel is doubly exponentially suppressed and therefore
we expect a marginal contribution to the effective annihilation
cross-section of Eq.~(\ref{reliccross}). In this sense, the link between
the relic density constraint and the strength of a collider signal is
much less direct.
This type of signal is already being analysed by
ATLAS~\cite{ATLAS:2013qla} and, although the background is smaller
than the one for a single pair of opposite-sign same-flavour leptons
plus $\cancel{E}_T$, it is still necessary to require a sizeable $\cancel{E}_T$
in order to suppress SM backgrounds, $\cancel{E}_T\gtrsim 50 $ GeV.
To obtain a sufficiently large $\cancel{E}_T$, one could again require an
additional ISR jet, and thus consider $ l^+l^-l'^+l'^-$ + jet + $\cancel{E}_T$.
The cross-section for this type of signal can be roughly estimated as
$\sigma(l^+l^-l'^+l'^{-}+j+\chi_1\bar{\chi}_1)\sim
(1/10)\times\sigma(l^{+}l^{-}+j+\chi_1\bar{\chi}_1)$.  Using the
parameter examples in Table~\ref{tab:num1}, we expect a non-negligible
signal for dark matter masses up to about 600 GeV.  We leave a proper
analysis of this interesting signal for future work.

\section{Conclusion}
\label{sec:conclusion}

We have examined a scenario in which the dark sector contains two
nearly degenerate particles: the DM candidate $\chi_1$, and a slightly
heavier state $\chi_2$.  In this scenario, the relic DM density is
determined by co-annihilation processes such as $\chi_1 + \overline{\chi_2}
\rightarrow SM $, while  direct and indirect detection processes are highly
suppressed, because they involve $\chi_1$ alone.  For standard
self-annihilating WIMPs, there is some tension between the size of the
couplings needed to obtain the correct relic abundance, and those
necessary to account for the non-observation of signals in direct
detection experiments.  The co-annihilation scenario eliminates this
incompatibility.

We described the interaction of $\chi_1$ and $\chi_2$ with SM
particles, by generalising the standard EFT description.  The
co-annihilation model offers interesting new collider signals: In
addition to the standard mono-jet + missing $E_T$ type of DM process, new
signals arise due to $\chi_1\overline{\chi_2}$ production followed by
the decay $\chi_2 \rightarrow \chi_1\,l^+l^-$ or $\chi_2 \rightarrow
\chi_1\,q \overline{q}$.  We have simulated signal and background for
the $l^+l^-$ + jet + missing $E_T$ process, for parameters which
correctly reproduce the DM relic density, and demonstrated that the
LHC has the potential to identify these signals with forthcoming data.
Of course, if dark matter is discovered at colliders, signals in
multiple channels would assist in uncovering its true nature, and the
new processes studied here would provide important complementary
information to the standard monojet type searches.

\section*{Acknowledgements}
We thank Tony Limosani, Martin White, and especially to Lei Wu and Brian Petersen
for helpful comments and advice.  We are also indebted to Heather
Logan for comments and for pointing out an important error in an
earlier version of this manuscript.  NFB, YC and ADM were supported by
the Australian Research Council.

\bibliography{DMCollider}

\begin{thebibliography}{59}
\expandafter\ifx\csname natexlab\endcsname\relax\def\natexlab#1{#1}\fi
\expandafter\ifx\csname bibnamefont\endcsname\relax
  \def\bibnamefont#1{#1}\fi
\expandafter\ifx\csname bibfnamefont\endcsname\relax
  \def\bibfnamefont#1{#1}\fi
\expandafter\ifx\csname citenamefont\endcsname\relax
  \def\citenamefont#1{#1}\fi
\expandafter\ifx\csname url\endcsname\relax
  \def\url#1{\texttt{#1}}\fi
\expandafter\ifx\csname urlprefix\endcsname\relax\def\urlprefix{URL }\fi
\providecommand{\bibinfo}[2]{#2}
\providecommand{\eprint}[2][]{\url{#2}}

\bibitem[{\citenamefont{Agnese et~al.}(2013)}]{Agnese:2013rvf}
\bibinfo{author}{\bibfnamefont{R.}~\bibnamefont{Agnese}} \bibnamefont{et~al.}
  (\bibinfo{collaboration}{CDMS Collaboration}),
  \bibinfo{journal}{Phys.Rev.Lett.} \textbf{\bibinfo{volume}{111}},
  \bibinfo{pages}{251301} (\bibinfo{year}{2013}), \eprint{1304.4279}.

\bibitem[{\citenamefont{Aalseth et~al.}(2013)}]{Aalseth:2012if}
\bibinfo{author}{\bibfnamefont{C.}~\bibnamefont{Aalseth}} \bibnamefont{et~al.}
  (\bibinfo{collaboration}{CoGeNT Collaboration}), \bibinfo{journal}{Phys.Rev.}
  \textbf{\bibinfo{volume}{D88}}, \bibinfo{pages}{012002}
  (\bibinfo{year}{2013}), \eprint{1208.5737}.

\bibitem[{\citenamefont{Angloher et~al.}(2012)\citenamefont{Angloher, Bauer,
  Bavykina, Bento, Bucci et~al.}}]{Angloher:2011uu}
\bibinfo{author}{\bibfnamefont{G.}~\bibnamefont{Angloher}},
  \bibinfo{author}{\bibfnamefont{M.}~\bibnamefont{Bauer}},
  \bibinfo{author}{\bibfnamefont{I.}~\bibnamefont{Bavykina}},
  \bibinfo{author}{\bibfnamefont{A.}~\bibnamefont{Bento}},
  \bibinfo{author}{\bibfnamefont{C.}~\bibnamefont{Bucci}},
  \bibnamefont{et~al.}, \bibinfo{journal}{Eur.Phys.J.}
  \textbf{\bibinfo{volume}{C72}}, \bibinfo{pages}{1971} (\bibinfo{year}{2012}),
  \eprint{1109.0702}.

\bibitem[{\citenamefont{Savage et~al.}(2009)\citenamefont{Savage, Gelmini,
  Gondolo, and Freese}}]{Savage:2008er}
\bibinfo{author}{\bibfnamefont{C.}~\bibnamefont{Savage}},
  \bibinfo{author}{\bibfnamefont{G.}~\bibnamefont{Gelmini}},
  \bibinfo{author}{\bibfnamefont{P.}~\bibnamefont{Gondolo}}, \bibnamefont{and}
  \bibinfo{author}{\bibfnamefont{K.}~\bibnamefont{Freese}},
  \bibinfo{journal}{JCAP} \textbf{\bibinfo{volume}{0904}}, \bibinfo{pages}{010}
  (\bibinfo{year}{2009}), \eprint{0808.3607}.

\bibitem[{\citenamefont{Ahmed et~al.}(2010)}]{Ahmed:2009zw}
\bibinfo{author}{\bibfnamefont{Z.}~\bibnamefont{Ahmed}} \bibnamefont{et~al.}
  (\bibinfo{collaboration}{CDMS-II Collaboration}), \bibinfo{journal}{Science}
  \textbf{\bibinfo{volume}{327}}, \bibinfo{pages}{1619} (\bibinfo{year}{2010}),
  \eprint{0912.3592}.

\bibitem[{\citenamefont{Ahmed et~al.}(2011)}]{Ahmed:2010wy}
\bibinfo{author}{\bibfnamefont{Z.}~\bibnamefont{Ahmed}} \bibnamefont{et~al.}
  (\bibinfo{collaboration}{CDMS-II Collaboration}),
  \bibinfo{journal}{Phys.Rev.Lett.} \textbf{\bibinfo{volume}{106}},
  \bibinfo{pages}{131302} (\bibinfo{year}{2011}), \eprint{1011.2482}.

\bibitem[{\citenamefont{Agnese et~al.}(2014)}]{Agnese:2013jaa}
\bibinfo{author}{\bibfnamefont{R.}~\bibnamefont{Agnese}} \bibnamefont{et~al.}
  (\bibinfo{collaboration}{SuperCDMSSoudan Collaboration}),
  \bibinfo{journal}{Phys.Rev.Lett.} \textbf{\bibinfo{volume}{112}},
  \bibinfo{pages}{041302} (\bibinfo{year}{2014}), \eprint{1309.3259}.

\bibitem[{\citenamefont{Akerib et~al.}(2013)}]{Akerib:2013tjd}
\bibinfo{author}{\bibfnamefont{D.}~\bibnamefont{Akerib}} \bibnamefont{et~al.}
  (\bibinfo{collaboration}{LUX Collaboration}) (\bibinfo{year}{2013}),
  \eprint{1310.8214}.

\bibitem[{\citenamefont{Angle et~al.}(2011)}]{Angle:2011th}
\bibinfo{author}{\bibfnamefont{J.}~\bibnamefont{Angle}} \bibnamefont{et~al.}
  (\bibinfo{collaboration}{XENON10 Collaboration}),
  \bibinfo{journal}{Phys.Rev.Lett.} \textbf{\bibinfo{volume}{107}},
  \bibinfo{pages}{051301} (\bibinfo{year}{2011}), \eprint{1104.3088}.

\bibitem[{\citenamefont{{The ATLAS collaboration}}(2012)}]{ATLAS:2012zim}
\bibinfo{author}{\bibnamefont{{The ATLAS collaboration}}}
  (\bibinfo{collaboration}{ATLAS Collaboration}) (\bibinfo{year}{2012}).

\bibitem[{\citenamefont{Aad et~al.}(2013{\natexlab{a}})}]{ATLAS:2012ky}
\bibinfo{author}{\bibfnamefont{G.}~\bibnamefont{Aad}} \bibnamefont{et~al.}
  (\bibinfo{collaboration}{ATLAS Collaboration}), \bibinfo{journal}{JHEP}
  \textbf{\bibinfo{volume}{1304}}, \bibinfo{pages}{075}
  (\bibinfo{year}{2013}{\natexlab{a}}), \eprint{1210.4491}.

\bibitem[{\citenamefont{Chatrchyan
  et~al.}(2012{\natexlab{a}})}]{Chatrchyan:2012tea}
\bibinfo{author}{\bibfnamefont{S.}~\bibnamefont{Chatrchyan}}
  \bibnamefont{et~al.} (\bibinfo{collaboration}{CMS Collaboration}),
  \bibinfo{journal}{Phys.Rev.Lett.} \textbf{\bibinfo{volume}{108}},
  \bibinfo{pages}{261803} (\bibinfo{year}{2012}{\natexlab{a}}),
  \eprint{1204.0821}.

\bibitem[{\citenamefont{Chatrchyan
  et~al.}(2012{\natexlab{b}})}]{Chatrchyan:2012me}
\bibinfo{author}{\bibfnamefont{S.}~\bibnamefont{Chatrchyan}}
  \bibnamefont{et~al.} (\bibinfo{collaboration}{CMS Collaboration}),
  \bibinfo{journal}{JHEP} \textbf{\bibinfo{volume}{1209}}, \bibinfo{pages}{094}
  (\bibinfo{year}{2012}{\natexlab{b}}), \eprint{1206.5663}.

\bibitem[{\citenamefont{Collaboration}(2013)}]{CMS:2013iea}
\bibinfo{author}{\bibfnamefont{C.}~\bibnamefont{Collaboration}}
  (\bibinfo{collaboration}{CMS Collaboration}) (\bibinfo{year}{2013}).

\bibitem[{\citenamefont{Adriani et~al.}(2009{\natexlab{a}})}]{Adriani:2008zr}
\bibinfo{author}{\bibfnamefont{O.}~\bibnamefont{Adriani}} \bibnamefont{et~al.}
  (\bibinfo{collaboration}{PAMELA Collaboration}), \bibinfo{journal}{Nature}
  \textbf{\bibinfo{volume}{458}}, \bibinfo{pages}{607}
  (\bibinfo{year}{2009}{\natexlab{a}}), \eprint{0810.4995}.

\bibitem[{\citenamefont{Adriani
  et~al.}(2010{\natexlab{a}})\citenamefont{Adriani, Barbarino, Bazilevskaya,
  Bellotti, Boezio et~al.}}]{Adriani:2010ib}
\bibinfo{author}{\bibfnamefont{O.}~\bibnamefont{Adriani}},
  \bibinfo{author}{\bibfnamefont{G.}~\bibnamefont{Barbarino}},
  \bibinfo{author}{\bibfnamefont{G.}~\bibnamefont{Bazilevskaya}},
  \bibinfo{author}{\bibfnamefont{R.}~\bibnamefont{Bellotti}},
  \bibinfo{author}{\bibfnamefont{M.}~\bibnamefont{Boezio}},
  \bibnamefont{et~al.}, \bibinfo{journal}{Astropart.Phys.}
  \textbf{\bibinfo{volume}{34}}, \bibinfo{pages}{1}
  (\bibinfo{year}{2010}{\natexlab{a}}), \eprint{1001.3522}.

\bibitem[{\citenamefont{Barwick et~al.}(1997)}]{Barwick:1997ig}
\bibinfo{author}{\bibfnamefont{S.}~\bibnamefont{Barwick}} \bibnamefont{et~al.}
  (\bibinfo{collaboration}{HEAT Collaboration}),
  \bibinfo{journal}{Astrophys.J.} \textbf{\bibinfo{volume}{482}},
  \bibinfo{pages}{L191} (\bibinfo{year}{1997}), \eprint{astro-ph/9703192}.

\bibitem[{\citenamefont{Aguilar et~al.}(2007)}]{Aguilar:2007yf}
\bibinfo{author}{\bibfnamefont{M.}~\bibnamefont{Aguilar}} \bibnamefont{et~al.}
  (\bibinfo{collaboration}{AMS-01 Collaboration}),
  \bibinfo{journal}{Phys.Lett.} \textbf{\bibinfo{volume}{B646}},
  \bibinfo{pages}{145} (\bibinfo{year}{2007}), \eprint{astro-ph/0703154}.

\bibitem[{\citenamefont{Ackermann et~al.}(2012)}]{FermiLAT:2011ab}
\bibinfo{author}{\bibfnamefont{M.}~\bibnamefont{Ackermann}}
  \bibnamefont{et~al.} (\bibinfo{collaboration}{Fermi LAT Collaboration}),
  \bibinfo{journal}{Phys.Rev.Lett.} \textbf{\bibinfo{volume}{108}},
  \bibinfo{pages}{011103} (\bibinfo{year}{2012}), \eprint{1109.0521}.

\bibitem[{\citenamefont{Adriani
  et~al.}(2009{\natexlab{b}})\citenamefont{Adriani, Barbarino, Bazilevskaya,
  Bellotti, Boezio et~al.}}]{Adriani:2008zq}
\bibinfo{author}{\bibfnamefont{O.}~\bibnamefont{Adriani}},
  \bibinfo{author}{\bibfnamefont{G.}~\bibnamefont{Barbarino}},
  \bibinfo{author}{\bibfnamefont{G.}~\bibnamefont{Bazilevskaya}},
  \bibinfo{author}{\bibfnamefont{R.}~\bibnamefont{Bellotti}},
  \bibinfo{author}{\bibfnamefont{M.}~\bibnamefont{Boezio}},
  \bibnamefont{et~al.}, \bibinfo{journal}{Phys.Rev.Lett.}
  \textbf{\bibinfo{volume}{102}}, \bibinfo{pages}{051101}
  (\bibinfo{year}{2009}{\natexlab{b}}), \eprint{0810.4994}.

\bibitem[{\citenamefont{Adriani et~al.}(2010{\natexlab{b}})}]{Adriani:2010rc}
\bibinfo{author}{\bibfnamefont{O.}~\bibnamefont{Adriani}} \bibnamefont{et~al.}
  (\bibinfo{collaboration}{PAMELA Collaboration}),
  \bibinfo{journal}{Phys.Rev.Lett.} \textbf{\bibinfo{volume}{105}},
  \bibinfo{pages}{121101} (\bibinfo{year}{2010}{\natexlab{b}}),
  \eprint{1007.0821}.

\bibitem[{\citenamefont{Chang et~al.}(2008)\citenamefont{Chang, Adams, Ahn,
  Bashindzhagyan, Christl et~al.}}]{Chang:2008aa}
\bibinfo{author}{\bibfnamefont{J.}~\bibnamefont{Chang}},
  \bibinfo{author}{\bibfnamefont{J.}~\bibnamefont{Adams}},
  \bibinfo{author}{\bibfnamefont{H.}~\bibnamefont{Ahn}},
  \bibinfo{author}{\bibfnamefont{G.}~\bibnamefont{Bashindzhagyan}},
  \bibinfo{author}{\bibfnamefont{M.}~\bibnamefont{Christl}},
  \bibnamefont{et~al.}, \bibinfo{journal}{Nature}
  \textbf{\bibinfo{volume}{456}}, \bibinfo{pages}{362} (\bibinfo{year}{2008}).

\bibitem[{\citenamefont{Aharonian et~al.}(2009)}]{Aharonian:2009ah}
\bibinfo{author}{\bibfnamefont{F.}~\bibnamefont{Aharonian}}
  \bibnamefont{et~al.} (\bibinfo{collaboration}{H.E.S.S. Collaboration}),
  \bibinfo{journal}{Astron.Astrophys.} \textbf{\bibinfo{volume}{508}},
  \bibinfo{pages}{561} (\bibinfo{year}{2009}), \eprint{0905.0105}.

\bibitem[{\citenamefont{Goodman et~al.}(2011)\citenamefont{Goodman, Ibe,
  Rajaraman, Shepherd, Tait et~al.}}]{Goodman:2010yf}
\bibinfo{author}{\bibfnamefont{J.}~\bibnamefont{Goodman}},
  \bibinfo{author}{\bibfnamefont{M.}~\bibnamefont{Ibe}},
  \bibinfo{author}{\bibfnamefont{A.}~\bibnamefont{Rajaraman}},
  \bibinfo{author}{\bibfnamefont{W.}~\bibnamefont{Shepherd}},
  \bibinfo{author}{\bibfnamefont{T.~M.} \bibnamefont{Tait}},
  \bibnamefont{et~al.}, \bibinfo{journal}{Phys.Lett.}
  \textbf{\bibinfo{volume}{B695}}, \bibinfo{pages}{185} (\bibinfo{year}{2011}),
  \eprint{1005.1286}.

\bibitem[{\citenamefont{Goodman et~al.}(2010)\citenamefont{Goodman, Ibe,
  Rajaraman, Shepherd, Tait et~al.}}]{Goodman:2010ku}
\bibinfo{author}{\bibfnamefont{J.}~\bibnamefont{Goodman}},
  \bibinfo{author}{\bibfnamefont{M.}~\bibnamefont{Ibe}},
  \bibinfo{author}{\bibfnamefont{A.}~\bibnamefont{Rajaraman}},
  \bibinfo{author}{\bibfnamefont{W.}~\bibnamefont{Shepherd}},
  \bibinfo{author}{\bibfnamefont{T.~M.} \bibnamefont{Tait}},
  \bibnamefont{et~al.}, \bibinfo{journal}{Phys.Rev.}
  \textbf{\bibinfo{volume}{D82}}, \bibinfo{pages}{116010}
  (\bibinfo{year}{2010}), \eprint{1008.1783}.

\bibitem[{\citenamefont{Dreiner et~al.}(2013)\citenamefont{Dreiner, Huck,
  Kramer, Schmeier, and Tattersall}}]{Dreiner:2012xm}
\bibinfo{author}{\bibfnamefont{H.}~\bibnamefont{Dreiner}},
  \bibinfo{author}{\bibfnamefont{M.}~\bibnamefont{Huck}},
  \bibinfo{author}{\bibfnamefont{M.}~\bibnamefont{Kramer}},
  \bibinfo{author}{\bibfnamefont{D.}~\bibnamefont{Schmeier}}, \bibnamefont{and}
  \bibinfo{author}{\bibfnamefont{J.}~\bibnamefont{Tattersall}},
  \bibinfo{journal}{Phys.Rev.} \textbf{\bibinfo{volume}{D87}},
  \bibinfo{pages}{075015} (\bibinfo{year}{2013}), \eprint{1211.2254}.

\bibitem[{\citenamefont{Beltran et~al.}(2010)\citenamefont{Beltran, Hooper,
  Kolb, Krusberg, and Tait}}]{Beltran:2010ww}
\bibinfo{author}{\bibfnamefont{M.}~\bibnamefont{Beltran}},
  \bibinfo{author}{\bibfnamefont{D.}~\bibnamefont{Hooper}},
  \bibinfo{author}{\bibfnamefont{E.~W.} \bibnamefont{Kolb}},
  \bibinfo{author}{\bibfnamefont{Z.~A.} \bibnamefont{Krusberg}},
  \bibnamefont{and} \bibinfo{author}{\bibfnamefont{T.~M.} \bibnamefont{Tait}},
  \bibinfo{journal}{JHEP} \textbf{\bibinfo{volume}{1009}}, \bibinfo{pages}{037}
  (\bibinfo{year}{2010}), \eprint{1002.4137}.

\bibitem[{\citenamefont{Fox et~al.}(2012)\citenamefont{Fox, Harnik, Kopp, and
  Tsai}}]{Fox:2011pm}
\bibinfo{author}{\bibfnamefont{P.~J.} \bibnamefont{Fox}},
  \bibinfo{author}{\bibfnamefont{R.}~\bibnamefont{Harnik}},
  \bibinfo{author}{\bibfnamefont{J.}~\bibnamefont{Kopp}}, \bibnamefont{and}
  \bibinfo{author}{\bibfnamefont{Y.}~\bibnamefont{Tsai}},
  \bibinfo{journal}{Phys.Rev.} \textbf{\bibinfo{volume}{D85}},
  \bibinfo{pages}{056011} (\bibinfo{year}{2012}), \eprint{1109.4398}.

\bibitem[{\citenamefont{Zheng et~al.}(2012)\citenamefont{Zheng, Yu, Shao, Bi,
  Li et~al.}}]{Zheng:2010js}
\bibinfo{author}{\bibfnamefont{J.-M.} \bibnamefont{Zheng}},
  \bibinfo{author}{\bibfnamefont{Z.-H.} \bibnamefont{Yu}},
  \bibinfo{author}{\bibfnamefont{J.-W.} \bibnamefont{Shao}},
  \bibinfo{author}{\bibfnamefont{X.-J.} \bibnamefont{Bi}},
  \bibinfo{author}{\bibfnamefont{Z.}~\bibnamefont{Li}}, \bibnamefont{et~al.},
  \bibinfo{journal}{Nucl.Phys.} \textbf{\bibinfo{volume}{B854}},
  \bibinfo{pages}{350} (\bibinfo{year}{2012}), \eprint{1012.2022}.

\bibitem[{\citenamefont{Buckley}(2011)}]{Buckley:2011kk}
\bibinfo{author}{\bibfnamefont{M.~R.} \bibnamefont{Buckley}},
  \bibinfo{journal}{Phys.Rev.} \textbf{\bibinfo{volume}{D84}},
  \bibinfo{pages}{043510} (\bibinfo{year}{2011}), \eprint{1104.1429}.

\bibitem[{\citenamefont{Busoni et~al.}(2013)\citenamefont{Busoni, De~Simone,
  Morgante, and Riotto}}]{Busoni:2013lha}
\bibinfo{author}{\bibfnamefont{G.}~\bibnamefont{Busoni}},
  \bibinfo{author}{\bibfnamefont{A.}~\bibnamefont{De~Simone}},
  \bibinfo{author}{\bibfnamefont{E.}~\bibnamefont{Morgante}}, \bibnamefont{and}
  \bibinfo{author}{\bibfnamefont{A.}~\bibnamefont{Riotto}}
  (\bibinfo{year}{2013}), \eprint{1307.2253}.

\bibitem[{\citenamefont{Griest and Seckel}(1991)}]{Griest:1990kh}
\bibinfo{author}{\bibfnamefont{K.}~\bibnamefont{Griest}} \bibnamefont{and}
  \bibinfo{author}{\bibfnamefont{D.}~\bibnamefont{Seckel}},
  \bibinfo{journal}{Phys.Rev.} \textbf{\bibinfo{volume}{D43}},
  \bibinfo{pages}{3191} (\bibinfo{year}{1991}).

\bibitem[{\citenamefont{Edsjo and Gondolo}(1997)}]{Edsjo:1997bg}
\bibinfo{author}{\bibfnamefont{J.}~\bibnamefont{Edsjo}} \bibnamefont{and}
  \bibinfo{author}{\bibfnamefont{P.}~\bibnamefont{Gondolo}},
  \bibinfo{journal}{Phys.Rev.} \textbf{\bibinfo{volume}{D56}},
  \bibinfo{pages}{1879} (\bibinfo{year}{1997}), \eprint{hep-ph/9704361}.

\bibitem[{\citenamefont{Cao et~al.}(2011)\citenamefont{Cao, Chen, Li, and
  Zhang}}]{Cao:2009uw}
\bibinfo{author}{\bibfnamefont{Q.-H.} \bibnamefont{Cao}},
  \bibinfo{author}{\bibfnamefont{C.-R.} \bibnamefont{Chen}},
  \bibinfo{author}{\bibfnamefont{C.~S.} \bibnamefont{Li}}, \bibnamefont{and}
  \bibinfo{author}{\bibfnamefont{H.}~\bibnamefont{Zhang}},
  \bibinfo{journal}{JHEP} \textbf{\bibinfo{volume}{1108}}, \bibinfo{pages}{018}
  (\bibinfo{year}{2011}), \eprint{0912.4511}.

\bibitem[{\citenamefont{Zhou et~al.}(2013)\citenamefont{Zhou, Berge, and
  Whiteson}}]{Zhou:2013fla}
\bibinfo{author}{\bibfnamefont{N.}~\bibnamefont{Zhou}},
  \bibinfo{author}{\bibfnamefont{D.}~\bibnamefont{Berge}}, \bibnamefont{and}
  \bibinfo{author}{\bibfnamefont{D.}~\bibnamefont{Whiteson}}
  (\bibinfo{year}{2013}), \eprint{1302.3619}.

\bibitem[{\citenamefont{Aad et~al.}(2013{\natexlab{b}})}]{Aad:2012fw}
\bibinfo{author}{\bibfnamefont{G.}~\bibnamefont{Aad}} \bibnamefont{et~al.}
  (\bibinfo{collaboration}{ATLAS Collaboration}),
  \bibinfo{journal}{Phys.Rev.Lett.} \textbf{\bibinfo{volume}{110}},
  \bibinfo{pages}{011802} (\bibinfo{year}{2013}{\natexlab{b}}),
  \eprint{1209.4625}.

\bibitem[{\citenamefont{Bell et~al.}(2012)\citenamefont{Bell, Dent, Galea,
  Jacques, Krauss et~al.}}]{Bell:2012rg}
\bibinfo{author}{\bibfnamefont{N.~F.} \bibnamefont{Bell}},
  \bibinfo{author}{\bibfnamefont{J.~B.} \bibnamefont{Dent}},
  \bibinfo{author}{\bibfnamefont{A.~J.} \bibnamefont{Galea}},
  \bibinfo{author}{\bibfnamefont{T.~D.} \bibnamefont{Jacques}},
  \bibinfo{author}{\bibfnamefont{L.~M.} \bibnamefont{Krauss}},
  \bibnamefont{et~al.}, \bibinfo{journal}{Phys.Rev.}
  \textbf{\bibinfo{volume}{D86}}, \bibinfo{pages}{096011}
  (\bibinfo{year}{2012}), \eprint{1209.0231}.

\bibitem[{\citenamefont{Carpenter et~al.}(2012)\citenamefont{Carpenter, Nelson,
  Shimmin, Tait, and Whiteson}}]{Carpenter:2012rg}
\bibinfo{author}{\bibfnamefont{L.~M.} \bibnamefont{Carpenter}},
  \bibinfo{author}{\bibfnamefont{A.}~\bibnamefont{Nelson}},
  \bibinfo{author}{\bibfnamefont{C.}~\bibnamefont{Shimmin}},
  \bibinfo{author}{\bibfnamefont{T.~M.} \bibnamefont{Tait}}, \bibnamefont{and}
  \bibinfo{author}{\bibfnamefont{D.}~\bibnamefont{Whiteson}}
  (\bibinfo{year}{2012}), \eprint{1212.3352}.

\bibitem[{\citenamefont{Bai and Tait}(2013)}]{Bai:2012xg}
\bibinfo{author}{\bibfnamefont{Y.}~\bibnamefont{Bai}} \bibnamefont{and}
  \bibinfo{author}{\bibfnamefont{T.~M.} \bibnamefont{Tait}},
  \bibinfo{journal}{Phys.Lett.} \textbf{\bibinfo{volume}{B723}},
  \bibinfo{pages}{384} (\bibinfo{year}{2013}), \eprint{1208.4361}.

\bibitem[{\citenamefont{Aad et~al.}(2014)}]{Aad:2013oja}
\bibinfo{author}{\bibfnamefont{G.}~\bibnamefont{Aad}} \bibnamefont{et~al.}
  (\bibinfo{collaboration}{ATLAS Collaboration}),
  \bibinfo{journal}{Phys.Rev.Lett.} \textbf{\bibinfo{volume}{112}},
  \bibinfo{pages}{041802} (\bibinfo{year}{2014}), \eprint{1309.4017}.

\bibitem[{\citenamefont{Tucker-Smith and Weiner}(2001)}]{TuckerSmith:2001hy}
\bibinfo{author}{\bibfnamefont{D.}~\bibnamefont{Tucker-Smith}}
  \bibnamefont{and} \bibinfo{author}{\bibfnamefont{N.}~\bibnamefont{Weiner}},
  \bibinfo{journal}{Phys.Rev.} \textbf{\bibinfo{volume}{D64}},
  \bibinfo{pages}{043502} (\bibinfo{year}{2001}), \eprint{hep-ph/0101138}.

\bibitem[{\citenamefont{Tucker-Smith and Weiner}(2005)}]{TuckerSmith:2004jv}
\bibinfo{author}{\bibfnamefont{D.}~\bibnamefont{Tucker-Smith}}
  \bibnamefont{and} \bibinfo{author}{\bibfnamefont{N.}~\bibnamefont{Weiner}},
  \bibinfo{journal}{Phys.Rev.} \textbf{\bibinfo{volume}{D72}},
  \bibinfo{pages}{063509} (\bibinfo{year}{2005}), \eprint{hep-ph/0402065}.

\bibitem[{\citenamefont{Bai and Tait}(2012)}]{Bai:2011jg}
\bibinfo{author}{\bibfnamefont{Y.}~\bibnamefont{Bai}} \bibnamefont{and}
  \bibinfo{author}{\bibfnamefont{T.~M.} \bibnamefont{Tait}},
  \bibinfo{journal}{Phys.Lett.} \textbf{\bibinfo{volume}{B710}},
  \bibinfo{pages}{335} (\bibinfo{year}{2012}), \eprint{1109.4144}.

\bibitem[{\citenamefont{Kolb and Turner}(1990)}]{Kolb:1990}
\bibinfo{author}{\bibfnamefont{E.~W.} \bibnamefont{Kolb}} \bibnamefont{and}
  \bibinfo{author}{\bibfnamefont{M.}~\bibnamefont{Turner}},
  \emph{\bibinfo{title}{The Early Universe}} (\bibinfo{publisher}{Westview
  Press}, \bibinfo{year}{1990}).

\bibitem[{\citenamefont{Christensen and Duhr}(2009)}]{Christensen:2008py}
\bibinfo{author}{\bibfnamefont{N.~D.} \bibnamefont{Christensen}}
  \bibnamefont{and} \bibinfo{author}{\bibfnamefont{C.}~\bibnamefont{Duhr}},
  \bibinfo{journal}{Comput.Phys.Commun.} \textbf{\bibinfo{volume}{180}},
  \bibinfo{pages}{1614} (\bibinfo{year}{2009}), \eprint{0806.4194}.

\bibitem[{\citenamefont{Belanger et~al.}(2007)\citenamefont{Belanger, Boudjema,
  Pukhov, and Semenov}}]{Belanger:2006is}
\bibinfo{author}{\bibfnamefont{G.}~\bibnamefont{Belanger}},
  \bibinfo{author}{\bibfnamefont{F.}~\bibnamefont{Boudjema}},
  \bibinfo{author}{\bibfnamefont{A.}~\bibnamefont{Pukhov}}, \bibnamefont{and}
  \bibinfo{author}{\bibfnamefont{A.}~\bibnamefont{Semenov}},
  \bibinfo{journal}{Comput.Phys.Commun.} \textbf{\bibinfo{volume}{176}},
  \bibinfo{pages}{367} (\bibinfo{year}{2007}), \eprint{hep-ph/0607059}.

\bibitem[{\citenamefont{Ade et~al.}(2013)}]{Ade:2013ktc}
\bibinfo{author}{\bibfnamefont{P.}~\bibnamefont{Ade}} \bibnamefont{et~al.}
  (\bibinfo{collaboration}{Planck Collaboration}) (\bibinfo{year}{2013}),
  \eprint{1303.5062}.

\bibitem[{\citenamefont{{The LEP Collaborations}}(2003)}]{LEP:2003aa}
\bibinfo{author}{\bibnamefont{{The LEP Collaborations}}}
  (\bibinfo{year}{2003}), \eprint{hep-ex/0312023}.

\bibitem[{\citenamefont{Buckley et~al.}(2011)\citenamefont{Buckley, Hooper,
  Kopp, and Neil}}]{Buckley:2011vc}
\bibinfo{author}{\bibfnamefont{M.~R.} \bibnamefont{Buckley}},
  \bibinfo{author}{\bibfnamefont{D.}~\bibnamefont{Hooper}},
  \bibinfo{author}{\bibfnamefont{J.}~\bibnamefont{Kopp}}, \bibnamefont{and}
  \bibinfo{author}{\bibfnamefont{E.}~\bibnamefont{Neil}},
  \bibinfo{journal}{Phys.Rev.} \textbf{\bibinfo{volume}{D83}},
  \bibinfo{pages}{115013} (\bibinfo{year}{2011}), \eprint{1103.6035}.

\bibitem[{\citenamefont{Dobrescu and Yu}(2013)}]{Dobrescu:2013cmh}
\bibinfo{author}{\bibfnamefont{B.~A.} \bibnamefont{Dobrescu}} \bibnamefont{and}
  \bibinfo{author}{\bibfnamefont{F.}~\bibnamefont{Yu}},
  \bibinfo{journal}{Phys.Rev.} \textbf{\bibinfo{volume}{D88}},
  \bibinfo{pages}{035021} (\bibinfo{year}{2013}), \eprint{1306.2629}.

\bibitem[{\citenamefont{Alwall et~al.}(2011)\citenamefont{Alwall, Herquet,
  Maltoni, Mattelaer, and Stelzer}}]{Alwall:2011uj}
\bibinfo{author}{\bibfnamefont{J.}~\bibnamefont{Alwall}},
  \bibinfo{author}{\bibfnamefont{M.}~\bibnamefont{Herquet}},
  \bibinfo{author}{\bibfnamefont{F.}~\bibnamefont{Maltoni}},
  \bibinfo{author}{\bibfnamefont{O.}~\bibnamefont{Mattelaer}},
  \bibnamefont{and} \bibinfo{author}{\bibfnamefont{T.}~\bibnamefont{Stelzer}},
  \bibinfo{journal}{JHEP} \textbf{\bibinfo{volume}{1106}}, \bibinfo{pages}{128}
  (\bibinfo{year}{2011}), \eprint{1106.0522}.

\bibitem[{\citenamefont{Sjostrand et~al.}(2008)\citenamefont{Sjostrand, Mrenna,
  and Skands}}]{Sjostrand:2007gs}
\bibinfo{author}{\bibfnamefont{T.}~\bibnamefont{Sjostrand}},
  \bibinfo{author}{\bibfnamefont{S.}~\bibnamefont{Mrenna}}, \bibnamefont{and}
  \bibinfo{author}{\bibfnamefont{P.~Z.} \bibnamefont{Skands}},
  \bibinfo{journal}{Comput.Phys.Commun.} \textbf{\bibinfo{volume}{178}},
  \bibinfo{pages}{852} (\bibinfo{year}{2008}), \eprint{0710.3820}.

\bibitem[{\citenamefont{de~Favereau et~al.}(2014)}]{deFavereau:2013fsa}
\bibinfo{author}{\bibfnamefont{J.}~\bibnamefont{de~Favereau}}
  \bibnamefont{et~al.} (\bibinfo{collaboration}{DELPHES 3}),
  \bibinfo{journal}{JHEP} \textbf{\bibinfo{volume}{1402}}, \bibinfo{pages}{057}
  (\bibinfo{year}{2014}), \eprint{1307.6346}.

\bibitem[{\citenamefont{Conte et~al.}(2013)\citenamefont{Conte, Fuks, and
  Serret}}]{Conte:2012fm}
\bibinfo{author}{\bibfnamefont{E.}~\bibnamefont{Conte}},
  \bibinfo{author}{\bibfnamefont{B.}~\bibnamefont{Fuks}}, \bibnamefont{and}
  \bibinfo{author}{\bibfnamefont{G.}~\bibnamefont{Serret}},
  \bibinfo{journal}{Comput.Phys.Commun.} \textbf{\bibinfo{volume}{184}},
  \bibinfo{pages}{222} (\bibinfo{year}{2013}), \eprint{1206.1599}.

\bibitem[{\citenamefont{{The ATLAS
  collaboration}}(2013{\natexlab{a}})}]{TheATLAScollaboration:2013oia}
\bibinfo{author}{\bibnamefont{{The ATLAS collaboration}}},
  \bibinfo{journal}{ATLAS-CONF-2013-082, ATLAS-COM-CONF-2013-075}
  (\bibinfo{year}{2013}{\natexlab{a}}).

\bibitem[{\citenamefont{{The CMS Collaboration}}(2013)}]{CMS-PAS-EXO-12-048}
\bibinfo{author}{\bibnamefont{{The CMS Collaboration}}},
  \bibinfo{journal}{CMS-PAS-EXO-12-048}  (\bibinfo{year}{2013}).

\bibitem[{\citenamefont{An et~al.}(2012)\citenamefont{An, Ji, and
  Wang}}]{An:2012va}
\bibinfo{author}{\bibfnamefont{H.}~\bibnamefont{An}},
  \bibinfo{author}{\bibfnamefont{X.}~\bibnamefont{Ji}}, \bibnamefont{and}
  \bibinfo{author}{\bibfnamefont{L.-T.} \bibnamefont{Wang}},
  \bibinfo{journal}{JHEP} \textbf{\bibinfo{volume}{1207}}, \bibinfo{pages}{182}
  (\bibinfo{year}{2012}), \eprint{1202.2894}.

\bibitem[{\citenamefont{An et~al.}(2013)\citenamefont{An, Huo, and
  Wang}}]{An:2012ue}
\bibinfo{author}{\bibfnamefont{H.}~\bibnamefont{An}},
  \bibinfo{author}{\bibfnamefont{R.}~\bibnamefont{Huo}}, \bibnamefont{and}
  \bibinfo{author}{\bibfnamefont{L.-T.} \bibnamefont{Wang}},
  \bibinfo{journal}{Phys.Dark Univ.} \textbf{\bibinfo{volume}{2}},
  \bibinfo{pages}{50} (\bibinfo{year}{2013}), \eprint{1212.2221}.

\bibitem[{\citenamefont{{The ATLAS
  collaboration}}(2013{\natexlab{b}})}]{ATLAS:2013qla}
\bibinfo{author}{\bibnamefont{{The ATLAS collaboration}}},
  \bibinfo{journal}{ATLAS-CONF-2013-036, ATLAS-COM-CONF-2013-041}
  (\bibinfo{year}{2013}{\natexlab{b}}).

\end{thebibliography}


\end{document}